\documentclass{emulateapj}
\usepackage{apjfonts}

\newcommand{\eqref}[1]{Equation (\ref{#1})}
\newcommand{\tabref}[1]{Table~\ref{#1}}
\newcommand{\figref}[1]{Figure~\ref{#1}}
\newcommand{\logten}[1]{\log_{10}{#1}}
\newcommand{\C}{^{12}{\rm C}}
\newcommand{\XC}{X_{12}}
\newcommand{\Ne}{^{22}{\rm Ne}}
\newcommand{\XNe}{X_{22}}
\shorttitle{SN~Ia dependence on $^{22}$Ne}

\begin{document}

\submitted{Accepted by the Astrophysical Journal June 21, 2009}
\title{Evaluating Systematic Dependencies of Type Ia Supernovae:\\
The Influence of Progenitor $^{22}$Ne Content on Dynamics}

\author{
Dean M. Townsley\altaffilmark{1},
Aaron P. Jackson\altaffilmark{2},
Alan C. Calder\altaffilmark{2,3},
David A. Chamulak\altaffilmark{4},\\
Edward F. Brown\altaffilmark{4,5},
and
F.~X. Timmes\altaffilmark{5,6}
}

\altaffiltext{1}{
Department of Astronomy / Steward Observatory,
The University of Arizona, Tucson, AZ; townsley@as.arizona.edu
}
\altaffiltext{2}{
Department of Physics \& Astronomy,
The State University of New York - Stony Brook, Stony Brook, NY
}
\altaffiltext{2}{
New York Center for Computational Sciences,
The State University of New York - Stony Brook, Stony Brook, NY
}
\altaffiltext{4}{
Department of Physics \& Astronomy,
Michigan State University, East Lansing, MI
}
\altaffiltext{5}{
The Joint Institute for Nuclear Astrophysics
}
\altaffiltext{6}{
School of Earth and Space Exploration,
Arizona State University, Tempe, AZ
}

\begin{abstract}

We present a theoretical framework for formal study of systematic effects in
Supernovae Type Ia (SN~Ia) that utilizes two-dimensional simulations to
implement a form of the deflagration-detonation transition (DDT) explosion
scenario.  The framework is developed from a randomized initial condition
that leads to a sample of simulated SN~Ia whose $^{56}$Ni masses have a
similar average and range to those observed, and have many other modestly
realistic features such the velocity extent of intermediate mass elements.
The intended purpose is to enable statistically well-defined studies of both
physical and theoretical parameters of the SN~Ia explosion simulation.
We present here a thorough description of the outcome of the SN~Ia explosions
produced by our current simulations.

A first application of this
framework is utilized to study the dependence of the SN~Ia on the $\Ne$
content, which is known to be directly influenced by the progenitor stellar
population's metallicity.   Our study is very specifically tailored to
measure how the $\Ne$ content influences the competition between the rise of
plumes of burned material and the expansion of the star before these plumes
reach DDT conditions.  This influence arises from the dependence of
the energy release, progenitor structure and laminar flame speed on
$\Ne$ content.  For this study, we explore these three effects for a fixed
carbon content and DDT density.  By setting the density at which
nucleosynthesis
takes place during the detonation phase of the explosion, the competition
between plume rise and stellar expansion controls the amount of material
in nuclear statistical equilibrium (NSE) and
therefore $^{56}$Ni produced.  Of particular interest is how this influence
of $\Ne$ content compares to the direct modification of the $^{56}$Ni mass
via the inherent neutron excess as discussed by Timmes, Brown \& Truran
(2003).  Although the outcome following from any particular ignition
condition can change dramatically with $\Ne$ content, with a sample of 20
ignition conditions we find that the systematic change in the expansion of
the star prior to detonation is not large enough to compete with the
dependence discussed by Timmes, Brown \& Truran (2003).  In fact, our results
show no statistically significant dependence of the pre-detonation expansion
on $\Ne$ content, pointing to the morphology of the ignition condition as
being the dominant dynamical driver of the $^{56}$Ni yield of the explosion.
However, variations in the DDT density, which were specifically excluded here,
are also expected to be important and
to depend systematically on $\Ne$ content.

\end{abstract}


%%%%%%%%%%%%%%%%%%%%%%%%%%%%%%%%%%%%%%%%%%%%%%%%%%%%%%%%%%%%%%%%%%%%%%%%
\section{Introduction}

Type Ia supernovae (SN~Ia) are bright stellar explosions spectroscopically
distinguished by strong silicon features near maximum light and a lack of 
hydrogen features~\citep{Fili97,hillebrandt.niemeyer:type}. 
Motivated broadly by the importance of SN~Ia for cosmological
studies~\citep{phillips:absolute,reispreskirs+96,albrecht_2006_aa}, contemporary observational
campaigns are gathering information about SN~Ia at an unprecedented
rate
(BAOSS, \citealt{Lietal96,Lietal01};
LOSS, \citealt{Trefetal97,Lietal01};
SNLS, \citealt{Astietal06};
CSP, \citealt{Hamuetal06};
Nearby SN Factory, \citealt{Copietal06};
Skymapper, \citealt{Kelletal07};
ESSENCE, \citealt{Woodetal07};
STRESS, \citealt{Bottetal08};
SDSS-II, \citealt{Holtetal08};
SXDS, \citealt{Furuetal08,Totaetal08};
PQ, \citealt{Djoretal08};
CfA, \citealt{Hicketal09a};
CRTS, \citealt{Draketal09};
PTF, \citealt{Kulketal09};
Pan-STARRS;
the La Silla SN Search)
and the future promises even more (the Dark Energy Survey, LSST, JDEM, see
\citealt{Howeetal09}).
Interesting systematics have been discovered and continue to be better
characterized.  All SN~Ia appear to burn a similar amount of total material,
but can differ widely in the amount of $^{56}$Ni produced, an effect that is
closely correlated with their brightness and decline time
\citep{Mazzetal07}.
Observations indicate that there are two populations
of SN~Ia that differ in the elapsed time between star formation
in the host galaxy and the explosion \citep{Scannapieco2005The-Type-Ia-Sup,Mannucci2005Two-populations}.
These two populations also appear to have a different average brightness,
with SN~Ia occurring in active star-forming galaxies appearing 
brighter~\citep{Howell2007Predicted-and-o}. Recent observations
find correlations suggesting that SNe Ia in galaxies whose populations 
have a characteristic age greater than 5 Gyr are ~1 mag fainter at 
maximum luminosity than those found in galaxies with younger 
populations, while progenitor metal abundance has a weaker influence
on peak luminosity~\citep{gallagheretal+08}.
The color properties of observed SNe~Ia have become an important
prospective source of systematic error for cosmic measurements
\citep{Hicketal09b}.  Alongside the finding that the scatter from the Hubble
law depends on host galaxy metallicity \citep{gallagheretal+08},  this
creates concern that corrections for extinction have become entangled with
uncharacterized intrinsic color variation.
It is in this context of extensive empirical searches for systematics that
our broad but still incomplete theoretical understanding of these events
must be pushed as far as possible, to attempt to predict
and understand systematic trends.

By invoking a straightforward counting argument based on the fact that the
number of protons and neutrons are approximately conserved in the explosion,
\citet{timmes.brown.ea:variations} argued that there should be a fairly
robust metallicity effect on the average $^{56}$Ni yield of SNe~Ia, and
therefore, potentially, their brightness.  Motivated by the simplicity of
this effect and the important implications for cosmological usage of SNe~Ia,
significant effort has gone into measuring such a metallicity dependence
observationally, but a clear effect has proven elusive
(\citealt{GallGarnetal05};
\citealt{gallagheretal+08}; \citealt{howelletal+09}).  Constraining
metallicity dependence alone is challenging for two reasons beyond the fact
that the effect does not appear to be large.
It is fairly difficult to measure accurate
metallicities for the parent stellar population, and even so there are strong
systematic problems with such measurements due to the mass-metallicity
relationship within galaxies \citep{gallagheretal+08}.  Additionally there is
an apparently much stronger dependence of mean brightness of SN~Ia on the age
of the parent stellar population
\citep{gallagheretal+08,howelletal+09}.

The most recent observational work leaves the situation still murky.
\citet{gallagheretal+08} were unsuccessful at finding a metallicity
dependence of the average brightness, but did find a such a dependence of the
Hubble residual (scatter from the Hubble law) obtained from light-curve 
fitting.  In contrast,
\citet{howelletal+09} found a slight dependence of average brightness on
metallicity and no similar Hubble residual issue using a different
light-curve fitting method.  However, the effect is difficult to separate
from the dependence on mean stellar age.
Given such uncertainty, it is important to
continue to evaluate the influence other aspects of the explosion might
have on the effect of the simple overall neutron excess outlined by
\citet{timmes.brown.ea:variations}.  The importance is further stressed
because it is clear that other effects, including the intrinsic random 
variation of otherwise similar SNe~Ia, can be even larger in magnitude.

The systematic effects of metallicity on the SN~Ia outcome have been the
topic of a number of theoretical studies with a variety of explosion models.
Several of these were accomplished in one dimension
\citep{HoefWheeThie98,Iwametal99,Hoefetal00,DomiHoefStra01}, and therefore
bear revisiting with multi-dimensional models.  In addition to changing the
overall yield of $^{56}$Ni, the initial neutron excess, as set by the
metallicity, is important for the composition of intermediate layers that
act as an important opacity source during the photospheric phase of the
supernova \citep{DomiHoefStra01}.  The challenges of understanding the SN~Ia
phenomenon with multi-dimensional numerical models have since somewhat
overshadowed this type of study of population systematics.
For example, one multi-dimensional study by \citet{RoepGiesetal06} studied
systematics using a less parameterized supernova model, but as a result was
forced to treat marginally successful deflagration models that do not
produce realistic explosions.  Additionally, the neutron excess was treated
only in post-processing, thus excluding sensitivity to dynamical effects like
those studied here.

The number of possible systematic parameters to consider, along with the
intrinsic scatter expected from implementation of realistic turbulence
characteristics, leads us to develop a theoretical framework for
formally evaluating systematic effects and their possible statistical
significance.  Systematic effects that require study include both
physical ones such composition and ignition density and purely theoretical
ones such as detonation condition and flame model parameters.  In addition to
the deflagration-detonation transition (DDT) explosion model itself, the
basic component of this framework is an initial condition that defines a
theoretical population or ensemble from which we can draw a sample of
supernovae and study how the sample as a whole responds to parameter changes,
giving a more complete and statistically quantifiable picture of their
systematic impact.

The ultimate goal of the study of SN~Ia systematics is to understand the 
dependence of properties of a SN~Ia population on 
characteristics of the parent stellar population.  Among other benefits,
this understanding would allow known characteristics of 
the stellar populations of galaxies,
such as their cosmic history, to be utilized in understanding systematic
trends that may appear in SN~Ia, or to better understand the chemical
evolution of galaxies.  Stellar populations are most basically characterized
by their age and metallicity, or more realistically, some mixture of or
distribution of these parameters.  There are likely to be several other
important secondary parameters such as binarity, or environment (e.g.
ionizing background) that may have enough influence on a stellar
population to be reflected in its SN~Ia population.
The dominance of stellar population age and metallicity
\citep{Scannapieco2005The-Type-Ia-Sup,Mannucci2005Two-populations,gallagheretal+08,howelletal+09},
or, alternatively, host galaxy morphology or color, in observational
investigations to date indicates that such secondary factors are probably
small by comparison.  However, until the age dependence is fully understood,
it is important to be mindful of possible additional contributions.

Our aim in the present paper is far more modest.  We begin by considering the
dependence of the supernova on the composition of the exploding WD, and
refine this exploration to only particular dynamical aspects in order to
allow a targeted, conclusive result.  The question that we seek to answer is:
does the change in the dynamics of the explosion due to a different $^{22}$Ne
abundance in the progenitor introduce a significant systematic effect in
addition to the neutron excess discussed by
\citet{timmes.brown.ea:variations}?
This highly targeted scope is motivated by a desire give a clear and
extensive description of our framework.
This is the first time that we are applying two-dimensional DDT simulations as
a method for generating a semi-realistic theoretical sample of supernovae and
studying the systematics of that sample.

The DDT model proved to be one of the most
successful of the one-dimensional SN~Ia models \citep[e.g.][]{HoefKhok96}.
However, it was never satisfying that both the deflagration
velocity and the DDT transition density were treated essentially as free
parameters.
The hope of multi-dimensional models is that burning propagation during the
deflagration phase, which was necessarily parameterized in one dimension,
can be calculated directly.  This would remove another free parameter and
lead to more reliable models.  Unfortunately the manifestation of buoyancy
instabilities in multi-dimensional models became a serious challenge.
Even modest asymmetries in the initial conditions of the deflagration led to
far too little expansion of the star by the time that a traditional DDT would
occur \citep{NiemHillWoos96,calder.ea:_offset_ignition_1,calder.ea:_offset_ignition_2,LivnAsidHoef05}.
This can be ameliorated with particular choices of ignition condition,
allowing the main desirable features of the 1-d models to also be obtained
from multi-dimensional models
\citep{GoloNiem05,GameKhokOran05,RoepNiem07,BravGarc08}.

The degree to which such a symmetric deflagration is physical is still a
matter of some debate.  The distribution of flame ignition regions in time
and space remains fairly uncertain \citep{WoosWunsKuhl04,RoepHilletal06} as
well as the degree to which turbulence-flame interaction after ignition can
influence the subsequent spread of the flame
\citep{RoepWoosHill07,Jordetal08}.  When placed alongside the physical
uncertainty as to whether or not a transition to detonation can actually
occur \citep{Niem99}, in a certain light the DDT scenario may simply be
too contrived.  However, so far the evidence is not sufficient to disprove
it, and it continues to provide one of the best prospects for matching the
observations of the typical SN~Ia.

We begin in Section \ref{sec:compositionsystematics} by discussing the
variety of physical effects through which composition can lead to systematic
variation of SN~Ia properties.  This forms the context for how and why we
limit the scope of this first study to dynamic effects only.  Following this,
in Section~\ref{sec:technical},
some improvements to the numerical model of the explosion are discussed and
the extensions necessary to treat detonation-flame interaction and the
presence of $\Ne$ are presented.  The burning model, flame speed, and mesh
refinement are each discussed.
In Section \ref{sec:framework}, we present the
ignition condition on which our ensemble of SNe~Ia is built and use it to
construct a framework for evaluating the significance of systematic effects.
The varieties of yield arising in the theoretical SN~Ia population are
discussed along with how the initial condition appears to control the outcome.
A metric for measuring the expansion prior to DDT is introduced
and calibrated to reflect the mass of nuclear statistical 
equilibrium (NSE) material synthesized.  Section
\ref{sec:2dddt} presents a detailed description of the outcome of a
representative two-dimensional simulation and how it generally compares with
observations.  Some points of the implementation of the explosion model that
necessitate referring to the details of the explosion are also described here.
Finally, in Section \ref{sec:ne22systematic} the framework is applied to measure
the effects of $\Ne$ content on the expansion of the star at the time of the
DDT.  We then summarize our conclusions and discuss
future work in Section \ref{sec:conclusions}.

%%%%%%%%%%%%%%%%%%%%%%%%%%%%%%%%%%%%%%%%%%%%%%%%%%%%%%%%%%%%%%%%%%
\section{Systematic Influences of Composition}
\label{sec:compositionsystematics}

We begin with an overview of the physical effects via which composition can
influence the process and outcome of a DDT explosion.  The most important
material constituents of the WD are $^{12}$C, $^{16}$O, and $^{22}$Ne, and we
will frame our discussion in terms of these.  Composition can influence the
explosion through changes in several physical processes and properties
involved in the explosion.  These include ignition density, DDT density,
energy release, flame speed, WD structure, and neutron excess.  Detailed
treatment of the first two of these, ignition density and DDT density, will
be deferred to future work.  The others will all be treated here, with the
last, neutron excess, taken as the baseline effect to which others should be
compared.  Several of these effects involve significant uncertainties, and it
is useful to highlight each in turn.

The compositional structure of the carbon-oxygen WD that explodes is
determined principally by the post-main sequence evolution of the star of
which it is a remnant \citep{DomiHoefStra01}.  The inner several tenths of a
$M_\odot$ are formed during the convective core helium burning phase, and the
layers outside this in shell burning on the asymptotic giant branch
\citep[and references therein]{Straetal03}.  During helium burning, the C, N,
and O which was present in the initial star is transformed into $^{22}$Ne
\citep{timmes.brown.ea:variations}, leading to a direct dependence on
metallicity of the parent stellar population.  In addition, $^{22}$Ne is
formed in the pre-explosion convective carbon burning core at a comparable
abundance \citep{PiroBild07,ChamBrowTimm07}.

{\it Ignition density} -- The ignition density characterizes when, as the
result of accretion, the WD core begins runaway heating due to carbon fusion
outpacing neutrino losses \citep{Nomo82_1,WoosWeav86}.  The central density
at flame ignition, when convection is insufficient to spread the heating
throughout the core, is slightly less (see e.g. \citealt{PiroBild07}), and
is also often called the ignition density, the meaning usually being
discernible from context.  The flame ignition density is generally near
$\sim 2\times 10^9$~g~cm$^{-3}$ for successful models of SN~Ia
\citep[e.g.][]{Nomoetal84,HoefKhok96}.
As we do here,
\citet{HoefWheeThie98,Hoefetal00} adopted a single value for the
ignition density for their studies of composition dependence.
Although we exclude it here, the variation of ignition density is expected to
be a significant effect.  Along with the energetic variations due to the
carbon content discussed below, it is likely an important contribution to the
observed dependence on stellar population age, which has proven to be
stronger than any metallicity dependence
\citep{gallagheretal+08,howelletal+09}.

The precise value of the ignition density
is sensitive to several factors, each of which has remarkable uncertainties.
One of the reasons we will simply fix the ignition density for this study is
the variety of uncertainties which must be considered if it is varied.
The energy generation rate depends on the $^{12}$C abundance in the core, as
set by the evolution of the star that formed the WD.  Authoritative
calculation of this abundance remains elusive due to its dependence on the
details of convection during late core helium burning
\citep[e.g.][]{Straetal03}.  The core temperature of the WD is also
important, and so the thermal history of the core, notably the accretion rate
and possibly properties of the helium flashes \citep{Nomoetal84}.
Either carbon composition or thermal state could lead to metallicity
dependencies that are closely involved with both the evolution of the parent
stellar population and the still very poorly understood
\citep{BranchLivioetal95} process of progenitor system formation.
There are also uncertainties in the screening enhancement of
nuclear reactions at these high densities
\citep{Gasques2005Nuclear-fusion-,YakoGasqetal06}, and in the reaction  
rates
themselves.  In particular, the $\mathrm{^{12}C + ^{12}C}$ reaction  
cross-section is not experimentally determined down to the low center- 
of-mass energies relevant for ignition in white dwarfs.  There is  
evidence, from heavy-ion fusion reactions, of ``hinderance''---a  
suppression of the astrophysical S-factor---at sub-barrier energies  
\citep{JianRehmetal07,GasqBrowetal07}.  In the case of $\mathrm{^{12}C  
+ ^{12}C}$, however, resonances are predicted in the energy range of  
interest \citep{MV72,PBA06}, which could significantly increase the  
cross section by as much as two orders of magnitude  
\citep{CoopSteiBrow09} at temperatures $\sim 5\times 10^8\,\mathrm{K}$.

{\it DDT density} -- We make the
presumption that there is a unique characteristic density at which there is a
transition from deflagration to detonation. In future work, this will be
extended to comparison of flame width and turbulent state
\citep{NiemWoos97,KhokOranWhee97,GoloNiem05}, giving a more physical
transition point.  While the details of this transition remain difficult to
fully quantify \citep{Aspdetal08,Woosetal09}, it is very likely that it will
have direct dependencies on composition because both the reactivity and the
flame width depend on both the $\C$ and $\Ne$ abundances.
Note that in this case, the important composition is that
in the outer portions of the star.  In addition to having a higher $\C$
abundance than the center \citep[e.g.][]{Hoefetal00} because it is the
product of shell burning, this region will have a lower $\Ne$ abundance
because it will not have participated in the core convection during which
$\Ne$ is enhanced in the convection zone \citep{PiroBild07,Chametal07}.
Given the outcome of this study, showing that dynamical factors have little
impact on the NSE yield, the influence on the DDT density is expected to be
the principal avenue, beyond inherited neutron excess, for dependence on
metallicity.  The laminar flame studies of \citet{Chametal07} predict a 10\%
reduction in the DDT density for a increase in the $\Ne$ abundance of 0.02.
From results found here, we estimate that this might correspond to a 3\%
decrease in NSE mass. This is about half as strong as the dependence from
inherited neutron excess.

{\it Energy release} -- The amount of energy release per unit mass can affect
both the global dynamics of the expansion and eventual disruption of the WD,
as well as the more local dynamics related to the buoyancy that drives the
acceleration of the burning front during the deflagration phase.  The
composition is essential for both these effects in two ways.  First, the
gross nuclear energy available per unit mass is mostly sensitive to the
abundance of $\C$.  Due to mixing during the smoldering phase, the abundance
over a broader region of the WD is involved in this case, as opposed to
precisely at the center as is the case for the ignition density above.  Thus,
the products of both the helium core and shell burning are important.
Secondarily, the $\Ne$ abundance, by changing the balance of protons and
neutrons, can influence the NSE state to which material will flash.  More
neutron rich material favors more tightly bound material and therefore
releases more energy \citep{Caldetal07}.

{\it Flame speed} -- The propagation speed of the subsonic burning front
propagated by thermal diffusion, the flame, is sensitive to both the $\C$ and
$\Ne$ abundances due to their effects on the energy release and the speed of
the early stages of the nuclear fusion \citep{timmes_1992_aa,ChamBrowTimm07}.  This laminar
propagation speed is likely to be much less important after the strong
turbulence that results from the buoyancy instability develops.  Turbulence
has the ability to add a nearly arbitrary amount of local area to the flame
surface, therefore making the burning effectively independent of the laminar
propagations speed \citep{NiemHill95}.  However, there is a period of time
early in the deflagration phase of the supernova where the interaction of the
laminar flame speed with the lower strength turbulent velocity field from the
core convection will be important for setting the morphology of the burned
region at
the point when strong buoyancy takes over \citep{ZingDurs07}.  In this work,
we find an important sensitivity to the assumed outcome of these earliest
stages.

{\it WD structure} -- The WD is supported by pressure of degenerate
electrons.  The neutron excess, and therefore the $\Ne$ abundance, sets the
number of baryons worth of weight each electron must support.  Thus a WD of
the same mass with a higher neutron excess will be more compact because it
will have fewer total electrons.  (Conversely at the same central density, a
star with more $\Ne$, and thus more neutrons, will be slightly less massive.)
This has a concomitant effect on the density distribution throughout the
star.  This is a fairly small effect \citep{HoefWheeThie98}, but important to
treat appropriately due to the marginally bound nature of a
near-Chandrasekhar mass WD.

{\it Neutron excess} --  In addition to the indirect effects mentioned
in relation to energy release above, the neutron excess, as set principally
by the $\Ne$ content, has a
direct influence over the final nucleosynthetic products, particularly the
amount of $^{56}$Ni.  \citet{timmes.brown.ea:variations} showed that the
decrease in the $^{56}$Ni produced in the explosion, absent other factors,
should be linearly proportional to the $\Ne$ content and therefore the
original metallicity of the stellar population.  The distribution of $\Ne$
will again be important, with that in the inner bulk of the star being
important for the gross yield, and that in surface layers for opacity sources
in those regions of the ejecta.

In order to bring the scope of our study within tractable limits, it is
necessary to either neglect or exclude some of these effects with
assumptions.  Most notably we would like to avoid for now the possibly quite
complex dependence of ignition density on composition.  This can naturally be
done by considering only a single value of $X_{\C}$, and assuming that we are
only studying the variation for progenitors that resulted from the same
formation
history.  Note that since metallicity changes main sequence lifetimes, for
example, this simplification may be a rather unnatural one in contrast to
comparing SN~Ia at the same stellar population age \citep{HoefWheeThie98}.
Our second simplifying assumption will be to neglect the dependence of the
DDT density on composition.  Exploring the effect of DDT density will be 
undertaken in immediate future work
along with the compositional inhomogeneity with which it is intertwined.
These two sets of assumptions leave us to study the dependence arising from
how changes in $\Ne$ abundance modify the energy release, flame speed, WD
structure and neutron excess. Because \citet{timmes.brown.ea:variations} have
provided an excellent discussion of the direct influence of neutron excess,
we focus mainly on how these other effects may modify the robust conclusion
they reached.

%%%%%%%%%%%%%%%%%%%%%%%%%%%%%%%%%%%%%%%%%%%%%%%%%%%%%%%%%%%%%%%%%%
\section{Improvement and Extension of Numerical Model}
\label{sec:technical}

The essential components of the numerical code used in this work were
presented in \citet{Townetal07} and work referenced therein, but we will give
a brief overview.  The overarching code is formed by the FLASH Eulerian
compressible hydrodynamics code \citep{Fryxetal00,calder.fryxell.ea:on} with
modules added for the nuclear burning which occurs in SNe~Ia.  FLASH uses a
high-order shock-capturing compressible hydrodynamics method, the piecewise
parabolic method \citep[PPM,][]{Colella1984The-Piecewise-P}, adapted to
treat a general equation of state \citep[EOS,][]{colella_1985_aa}.  We use
FLASH's tabulated fully ionized electron-ion plasma EOS
\citep{timmes.swesty:accuracy,Fryxetal00}. FLASH applies this hydrodynamics
method on an adaptively refined, tree-structured, non-moving Eulerian grid.
We make extensive usage of this adaptive mesh refinement (AMR) capability,
using different resolutions for burning fronts (4 km), the initial
hydrostatic star (16 km), and the region of negligible density initially
outside the star (as coarse as thousands of km).  Extensive detail on our
refinement scheme and tests of resolution are given in
section~\ref{sec:refinement} below.

The nuclear burning processes, beginning with carbon fusion, and extending to
electron capture in material in NSE are
implemented with a nuclear energetics model described by \citet{Townetal07},
and here in section \ref{sec:burning}, and calibrated to reproduce the
features of nuclear processes that occur in SN~Ia as calculated using
hundreds of nuclides \citep{Caldetal07}.  This burning model is used for both
subsonic (deflagration) and supersonic (detonation) burning fronts.  Because
the flame physics that governs the propagation speed of the subsonic burning
front is unresolved at 4~km \citep{Chametal07}, we propagate this front by
coupling our nuclear energetics to an artificial, resolved reaction front
given by the advection-reaction-diffusion (ARD or ADR) equation
\citep{Khok95,VladWeirRyzh06}.  A variety of special measures are necessary
to ensure this coupling is acoustically quiet and stable, and therefore
appropriate for simulating the buoyant instability of the burning front
\citep{Townetal07}.  Detonation fronts are handled somewhat naturally by the
shock-capturing features of the hydrodynamics scheme \citep{Meaketal09}.  The
only common components between our code and that of \citet{Plew07} are those
publicly available as components of FLASH, which excludes all components
treating the nuclear burning; differences are discussed in
\citet{Townetal07}.

In this section we discuss various changes made to improve the overall
consistency of the numerical modeling as well as the extensions necessary for
the burning model and flame speed treatments to include an arbitrary
$^{22}$Ne content.  Changes to the burning model, flame speed and refinement
are discussed in turn, with separate subsections for improvements and
extensions.  The improvements include a backwards-differenced time
integration method for the nuclear energetics model, more well-defined
methods for defining the density used to compute the flame speed and the
Atwood number used to calculate the buoyancy-compensated ("turbulent") flame
speed, as well as simplified and better tested mesh refinement prescriptions.
Extensions include modifications to the coupling of the RD
front and nuclear energetics in order to better account for the interaction
of the detonation front with the artificially thickened sub-sonic burning
front, and inclusion of the dependence of the laminar flame speed on
composition.

\subsection{Burning Model}
\label{sec:burning}

Although the model of nuclear energy release used here is functionally
identical to that of \cite{Townetal07}, it is useful to restate it in a way
that makes adjusting the abundance of the fuel easier and that is more
amenable to backwards differencing or sub-stepping in time.  This will also
provide the context in which we can present our prescriptions for how the
detonation interacts with the artificially broad flame front.

\subsubsection{Simplified Dynamical Equations}

The first step
is to abstract the properties of the fuel and the
ashes of carbon burning into adjustable parameters.  The
properties of interest are the number of electrons per baryon, $Y_e$, the
number of fluid ions per baryon, $Y_{\rm ion}$, and the average nuclear
binding energy per baryon, $\bar q$.  These can be constructed from mass
fractions $X_i$ per
\begin{eqnarray}
Y_e &=& \sum_i \frac{Z_i}{A_i} X_i,\\
Y_{\rm ion} &=& \sum_i \frac{1}{A_i}X_i,\\
\bar q &=& \sum_i \frac{E_{b,i}}{A_i}X_i
\end{eqnarray}
where $Z_i$ is the nuclear charge, $A_i$ is the atomic mass number (number of
baryons), and $E_{b,i} = (Z_im_p-N_im_n-m_i)c^2$ is the nuclear binding
energy, where $N_i=A_i-Z_i$ and $m_p$, $m_n$, and $m_i$ are the rest masses
of the proton, neutron and nucleus of nuclide $i$.
Note that because the quantities $Y_e$, $Y_{\rm ion}$ and $\bar q$
are "per baryon", it is more clear to think of the density
on the grid as representing the baryon number density.  The advantage being
that, unlike mass, this is actually a conserved quantity, making it possible
to calculate the overall change in rest mass energy in a well-defined way.
Baryon number divided by Avogadro's number makes an extremely good
approximation for the mass in grams, and this correspondence will be used so
extensively that in situations where the difference is of no consequence, we
will almost always refer to the mass density instead.  In this interpretation
the $X_i$ are actually baryon number fractions and also become conserved
quantities in the absence of transformations.

We begin by defining the properties of "pure" fuel or carbon-burning ashes
as $Y_{e,f},Y_{{\rm ion},f},\bar q_f$ and $Y_{e,a},Y_{{\rm ion},a},\bar q_a$
respectively.  As before, we also define three progress variables, which
measure the progress of the burning through various stages.  These are
$\phi_{fa}$, for processing of fuel to carbon-burning ashes, $\phi_{aq}$, for
processing of these ashes to Nuclear Statistical Quasi-Equilibrium (NSQE),
and finally $\phi_{qN}$ for relaxation of NSQE to full nuclear statistical
equilibrium (NSE).  Finally we will define the properties of the NSQE+NSE
material, $\delta Y_{e,qN},\delta Y_{{\rm ion},qN},\delta \bar q_{qN}$ in
such a way that the bulk properties are
\begin{eqnarray}
Y_e &=& (1-\phi_{fa})Y_{e,f} + (\phi_{fa}-\phi_{aq})Y_{e,a} +
\delta Y_{e,qN} \\
Y_{\rm ion} &=& (1-\phi_{fa})Y_{{\rm ion},f} +(\phi_{fa}-\phi_{aq})Y_{{\rm
ion},a} + \delta Y_{{\rm ion},qN}\\
\label{eq:qbar}
\bar q &=& (1-\phi_{fa})\bar q_f + (\phi_{fa}-\phi_{aq})\bar q_a
+ \delta \bar q_{qN}\ .
\end{eqnarray}
The usage of, for example, $\delta \bar q_{qN}$ rather than $\bar q_{qN}$, as
was done in \citet{Townetal07}, is to avoid issues of how the evolution of the
NSE material should be treated when $\phi_{qN}$ is very small.

As discussed in \citet{Townetal07} it is possible to calculate a "final" NSE
state by calculating the endpoint of an isobaric or isochoric burning based
on the instantaneous local conditions.  If we denote this state by "NSE" and
also use the NSQE and NSE timescales parameterized based on the expected
temperature of this final state, $\tau_{NSQE}$ and $\tau_{NSE}$
\citep{Caldetal07},
 we can posit Lagrangian
source terms.
\begin{eqnarray}
\label{eq:dphifa}
\frac{D\phi_{fa}}{Dt}
&=&
	\max(0,\dot \phi_{\rm RD}) + \dot\phi_{\rm CC}\\
\frac{D\phi_{aq}}{Dt}
&=&
	\frac{\phi_{fa}-\phi_{aq}}{\tau_{NSQE}}\\
\frac{D\phi_{qN}}{Dt}
&=&
	\frac{\phi_{aq}-\phi_{qN}}{\tau_{NSE}}\\
\frac{D(\delta \bar q_{qN})}{Dt}
&=&
	\frac{D\phi_{aq}}{Dt}\bar q_{NSE} +
	\frac{\phi_{aq}\bar q_{NSE}-\delta \bar q_{qN}}{\tau_{NSQE}}\\
\frac{D(\delta Y_{e,N})}{Dt}
&=&
	\frac{D\phi_{qN}}{Dt}Y_{e,a}+\phi_{qN}\dot Y_{e,NSE}\\
\frac{D(\delta Y_{{\rm ion},qN})}{Dt}
&=&
	\frac{D\phi_{aq}}{Dt}\tilde Y_{{\rm ion},q}
	+ \frac{1}{\tau_{NSQE}}\bigl[
	(\phi_{aq}-\phi_{qN})\tilde Y_{{\rm ion},q}
\nonumber\\
&& \quad\quad\quad\quad\mbox{}
	+\phi_{qN}Y_{{\rm ion},NSE}-\delta Y_{{\rm ion},qN}\bigr].
\label{eq:dyiqn}
\end{eqnarray}
Here $\dot \phi_{\rm RD}$ is the contribution from the reaction-diffusion
model for the burning front propagation \citep{Townetal07},
\begin{equation}
\dot \phi_{\rm RD} = \kappa \nabla^2  \phi_{\rm RD} +
\frac{1}{\tau}R(\phi_{\rm RD})
\end{equation}
where $\phi_{\rm RD}$ is the progress variable for the (A)RD front
and the coefficients $\kappa$ and $\tau$ are determined from the prescribed
propagation speed of the front, $S$, and its desired width.
$R(\phi)=f(\phi-\epsilon_0)(1-\phi+\epsilon_1)/4$, with tunable parameters
$f$, and $0<\epsilon_0,\epsilon_1\ll1$, provides a stable and acoustically
quiet reaction front propagation as
discussed in \citet{Townetal07}.  Thermal fusion is included via the
reaction term
\begin{equation}
\dot\phi_{\rm CC} = \rho X_{{\rm ^{12}C},f}(1-\phi_{fa})^2\frac{1}{12}
N_{\rm A} \langle\sigma v\rangle_{\rm C+C}(T)
\end{equation}
where $N_A\langle \sigma v\rangle_{\rm C+C}(T)$ is the carbon-carbon fusion rate from
\citet{CaugFowl88}, $\rho$ is the local density, and $X_{^{12}{\rm C},f}$ is
the carbon mass fraction in pure fuel.
While this rate is subject to several important uncertainties as described
in section~\ref{sec:compositionsystematics} above, the features of the
detonation are fairly insensitive to its precise value, depending mainly on
the overall energy release.
Some additional prescriptions for
treating the temperature used to calculate this rate are required in regions
where the RD front is active, $\phi_{\rm RD} >0$, which will be discussed below.
In order to accurately propagate the detonation, $\dot\phi_{\rm CC}$ is set
to zero inside shocks \citep{FryxMuelArne89}.
Finally, $\tilde Y_{{\rm ion},q}$ is a very rough estimate of the ion
abundance for NSQE material given by
\begin{equation}
\tilde Y_{{\rm ion},q} = 
\max\left[\frac{1}{28},\min\left(\frac14,
\frac{Y_{\rm ion,NSE}-\frac{1}{56}}{\frac14-\frac{1}{56}}\frac14
+\frac{\frac14-Y_{\rm
ion,NSE}}{\frac14-\frac{1}{56}}\frac{1}{28}\right)\right].
\end{equation}
Conservation of mass energy yields the evolution of the mass-specific energy
due to these nuclear processes,
\begin{equation}
\label{eq:de}
\frac{De}{Dt}=
N_A\frac{D\bar q}{Dt}
- \phi_{qN}[\dot Y_{e,NSE}N_A(m_p+m_e-m_n)c^2 +\dot e_{\nu,NSE}],
\end{equation}
where $D\bar q/Dt$ is obtained using (\ref{eq:qbar}) and the above source
terms, $m_p$, $m_n$, and $m_e$ are the proton, neutron and electron masses
respectively, $c$ is the speed of light, and $\dot e_{\nu,NSE}$ is the energy
loss rate to neutrinos due to weak processes in the predicted NSE state
\citep{Caldetal07,SeitTownetal09}.

All the dynamical quantities in equations (\ref{eq:dphifa})-(\ref{eq:dyiqn})
are mass-specific, or more appropriately baryon-specific, quantities.  Their
full hydrodynamic evolution is thus given, for a representative quantity,
$q$, by
\begin{equation}
\frac{\partial (q\rho)}{\partial t}
= \left[-\nabla\cdot(q\rho\vec v)\right]_H + \left[\rho\frac{Dq}{Dt}\right]_B
\end{equation}
where $\vec v$ is the velocity field.  The portions indicated as $H$ and $B$
are treated in an operator split fashion, each acting consecutively, with $H$
being the conservative hydrodynamic operator implemented with Piecewise Parabolic
Method (PPM)~\citep{Colella1984The-Piecewise-P} and $B$ the
burning scheme described above.  Because the
$B$ operator also does not change $\rho$, equations
(\ref{eq:dphifa})-(\ref{eq:dyiqn}) reduce to simple time derivatives after
operator splitting.
For the application of the $B$ operator, we further assume that the
temperature used in the calculation of $\dot\phi_{\rm CC}$ as well as the
timescales and properties of the NSE final state are constant, so that
equations (\ref{eq:dphifa})-(\ref{eq:dyiqn}) can be backwards-differenced and
solved algebraically.  This is intended to increase the stability of the
treatment, particularly when $\tau_{\rm NSE}$ is similar to the size of a
hydrodynamic time step.  While the assumption that $T$ is constant is not a
good one for the thermal term, this is likely to have fairly little impact as
this term is largely only active in propagating the detonation, for which the
burning length is unresolved.
The NSE final state prediction should only vary slowly with time, even in
burning fronts, since it depends on the combination $e-N_A\bar q$ or
$e-PN_A\bar q/\rho$ for the isochoric and isobaric predictions respectively.
The former combination only varies on the hydrodynamic timescale and the
timescale of weak processes, and the latter should vary fairly slowly due to
the subsonic propagation of the RD front.

\subsubsection{Detonation-Flame Interaction}

The RD front that we utilize to model the propagation of the subsonic
burning front is several computational zones thick in order to enable its stable
propagation.  This presents several complications when it is desirable to
also treat thermally activated burning that is not related to the
flame front.  At high densities, the physical flame is very
thin, and as a result, there is very little partially burned material.  Thus,
in this density regime, partially burned material on the grid represents
regions where burned and unburned material are well separated, but the
interface is not resolved.  In addition, this means that the temperature of
the zone is not representative of the temperature in the unburned material,
which should be used in the calculation of $\dot\phi_{\rm CC}$.
In previous work \citep{Meaketal09}, the thermally activated term,
$\dot\phi_{\rm CC}$ in equation (\ref{eq:dphifa}) was suppressed (set to zero)
within the RD front, where $\phi_{\rm RD}$ exceeded some small threshold.  This is
not too bad an approximation in the case of the gravitationally confined
detonation (GCD) scenario because only a small portion of the star is burned
via the deflagration, so that an even smaller portion is left in an
unrealistic partially burned state.  However, such a prescription would leave
a large amount of partially burned material in a simulation in the DDT
scenario.

The principal intent of allowing a thermal contribution within the RD front
is to allow the detonation to consume nearly all of the partially burned
material as it encounters the RD front, as it would if encountering a thin
flame.  In order to accomplish this, two measures are applied.  First,
$\dot\phi_{\rm CC}$ is suppressed where $\phi_{\rm RD}
> 10^{-6}$ unless $\phi_{fa} -\phi_{\rm RD} > 0.1$ within one flame width
(4 zone widths) of a given cell.
This prevents spurious thermal burning.  Second, where
$\dot\phi_{\rm CC}$ is enabled within the RD front, an estimate of the
temperature of the unburned material is used in place of the local
temperature in the calculation of $\langle\sigma v\rangle_{\rm C+C}(T)$.
This temperature estimate
is obtained by estimating the properties the local material would have in the
absence of the RD-front based fuel consumption.  The fraction burned due to
thermal processes is $X_{b,\rm th}=\phi_{fa}-\phi_{\rm RD}$.
The properties in the absence of the RD front would then be 
\begin{eqnarray}
\bar q_{\rm th} &=& (1-X_{b,\rm th})\bar q_f + X_{b,\rm th}\bar q_b\\
Y_{{\rm ion},\rm th} &=& (1-X_{b,t})Y_{{\rm ion},\rm th} + X_{b,\rm th}Y_{{\rm ion},b}
\end{eqnarray}
 where we estimate
\begin{eqnarray}
\bar q_b &=& \frac{\bar q - (1-\phi_{fa})\bar q_f}{\phi_{fa}}\\
Y_{{\rm ion},b} &=& \frac{Y_{\rm ion} - (1-\phi_{fa})Y_{{\rm
ion},f}}{\phi_{fa}} \ .
\end{eqnarray}
Then the temperature estimate for the thermally activated burning, $T_{\rm
th}$, is found by evaluating the equation of state for the local density,
$Y_e$, $Y_{{\rm ion},\rm th}$ and mass-specific internal energy
given by $e_{\rm th} = e-(\bar q-\bar q_{\rm th})$.
With a sufficiently accurate and robust estimate of the temperature, the
first condition, suppressing $\dot\phi_{\rm CC}$ in the RD front except in
proximity to other burning, might prove unnecessary.  This was not the case
in tests we have run so far.  The above
prescriptions appear sufficient for
the purposes of this study.

\subsection{Flame Speed}
\label{sec:flamespeed}

The improvements to the flame speed treatment are intended only to improve
numerical consistency by obtaining a more well-defined and stable value for
the front propagation speed.  The improvements are to estimate the unburned
density and to use a tabulation of the Atwood number instead of a crude
estimate.  In doing this, we also develop methodology for testing the
accuracy of tabulations to ensure that the tabulation grid is sufficiently
fine.  Following this, we present details of our extension of the computation
of the laminar flame speed to account for dependence on fuel composition.
This is one of the physical effects that our study of $^{22}$Ne systematics
is evaluating.  The tabulation utilized for the laminar flame speed is also
evaluated for accuracy.

\subsubsection{Consistency Improvements of Flame Speed Treatment}

To be consistent with the burning model, the input flame front speed is 
a function of the local chemical composition of fuel and the density of that
fuel. We 
define the flame front speed to be the speed of the carbon burning front 
with respect to the fuel. We parameterize the chemical composition of the 
fuel by $\C$ and $\Ne$ mass fractions, with the remaining material being
$^{16}$O. In order to prevent the flame from being torn apart by turbulence
induced by the Rayleigh-Taylor (RT) instability, we introduce a minimum flame
speed,
\begin{equation}\label{fspd_min}
S_{{\rm min}} = 0.5\sqrt{Agm\Delta} {\rm ,}
\end{equation}
where $A$ is the Atwood number for carbon burning, $g$ is the gravitational
acceleration, $\Delta$ is the grid resolution, and $m$ is a calibrated
parameter determined to be $m \sim 0.04\textrm{--}0.06$~\citep{TownAsidetal09}. In this work, 
we use $m=0.04$.
The Atwood number and gravitational acceleration describe the strength of the
RT-induced turbulence.  The resulting speed used to set the propagation of
the ADR front is
\begin{equation}\label{fspd}
S = {\rm max}\left( S_{\rm min}, S_{\rm lam} \right ){\rm ,}
\end{equation}
where $S_{\rm lam}$ is the speed of the laminar flame front.  $S$ is then
rolled off to zero for densities below $10^7$~g~cm$^{-3}$.
In reality, the RT-induced turbulence wrinkles the flame, increasing the
surface area of the burning front and accordingly increasing the burning
rate. The minimum flame speed for our model thus serves to compensate for
this boosted burning rate~\citep{Khok95}.  By construction, this
prescription does not properly account for the interaction of isolated
turbulence with the flame surface.  Thus while the integrity of a given plume
is maintained, the interaction of turbulence created by one plume with others
is not necessarily captured accurately.  We perform resolution studies and
vary the parameter $m$ to check for precisely these issues (Section
\ref{sec:resstudy}) and find that they do not appear to adversely
affect the principal metric measured in this study, the expansion prior to
DDT.  Inclusion of more direct models of flame-turbulence interaction
\citep{Colietal00,Schmetal06a} is planned for future work.

Both $S_{\rm lam}$ and $A$ are best characterized by a dependency on the
composition and density of the unburned fuel.  This presents a difficulty
when $S$ must be constructed in a partially burned cell, where the density
and composition has changed from that of the unburned material due to burning
and energy release.  For the composition, we simply store separate mass
scalars that represent the initial composition: the mass fractions of
material that was initially in the form of $\C$ and
$\Ne$.  While the density varies across the subsonic burning front, the
pressure is approximately constant.  We therefore use the local pressure,
$P$, to estimate the density of the unburned state.  This is greatly
simplified by the fact that the unburned material in the WD is at a high
degree of degeneracy, so that the temperature of the unburned state is not
too important and simple forms for the pressure-density relation of a
degenerate gas can be used \citep{hankaw1994}.  Our estimate for the unburned
density is
\begin{equation}
\label{eq:rhou}
\rho_u = 0.92 \times \frac{1}{Y_e} \sqrt{
   \left(\frac{P}{1.243\times 10^{15}}\right)^{3/2} + 
   \left(\frac{P}{1.004\times 10^{13}}\right)^{6/5} } {\rm ,}
\end{equation}
where $P$ is the pressure in cgs units (erg~cm$^{-3}$), $Y_e$ is the electron
fraction, and 0.92 is adjusted to provide a good fit to the pressure-density
relation in the initial star.  The difference between this estimated density
based on the local pressure and the actual density is no more than 6\% in the
initial star for $\rho> 2~\times~10^5$~g~cm$^{-3}$ and varies smoothly from
an overestimate at low pressures to an underestimate at high pressures.

The Atwood number used in the calculation of the minimum flame speed
is determined based on the local fuel composition and the resulting energy
release from carbon burning.
Given these parameters, we estimate the 
density of the ash using the Rankine-Hugoniot jump condition across the 
flame front~\citep{VladWeirRyzh06}. For computational efficiency, we 
calculate the Atwood number at the beginning of the simulation
for twenty-one equidistant log densities 
between 6 and 9.6 and ten equidistant carbon mass fractions between 0.3 
and 1.0 using a representative neon mass fraction appropriate for the
simulation.  Because the
Atwood number changes less than $0.01\%$ over the relevant range of $\XNe$,
we reduce the dimensionality of the interpolation and memory requirements by
introducing a characteristic neon mass fraction, which for this study is equal
to the global neon mass fraction. During a 
simulation, the code performs a bi-linear interpolation of the Atwood number
given the local initial mass fraction of $\C$ and the estimated unburned
density from Eq. (\ref{eq:rhou}).

\begin{figure}[t]
\includegraphics[width=\columnwidth]{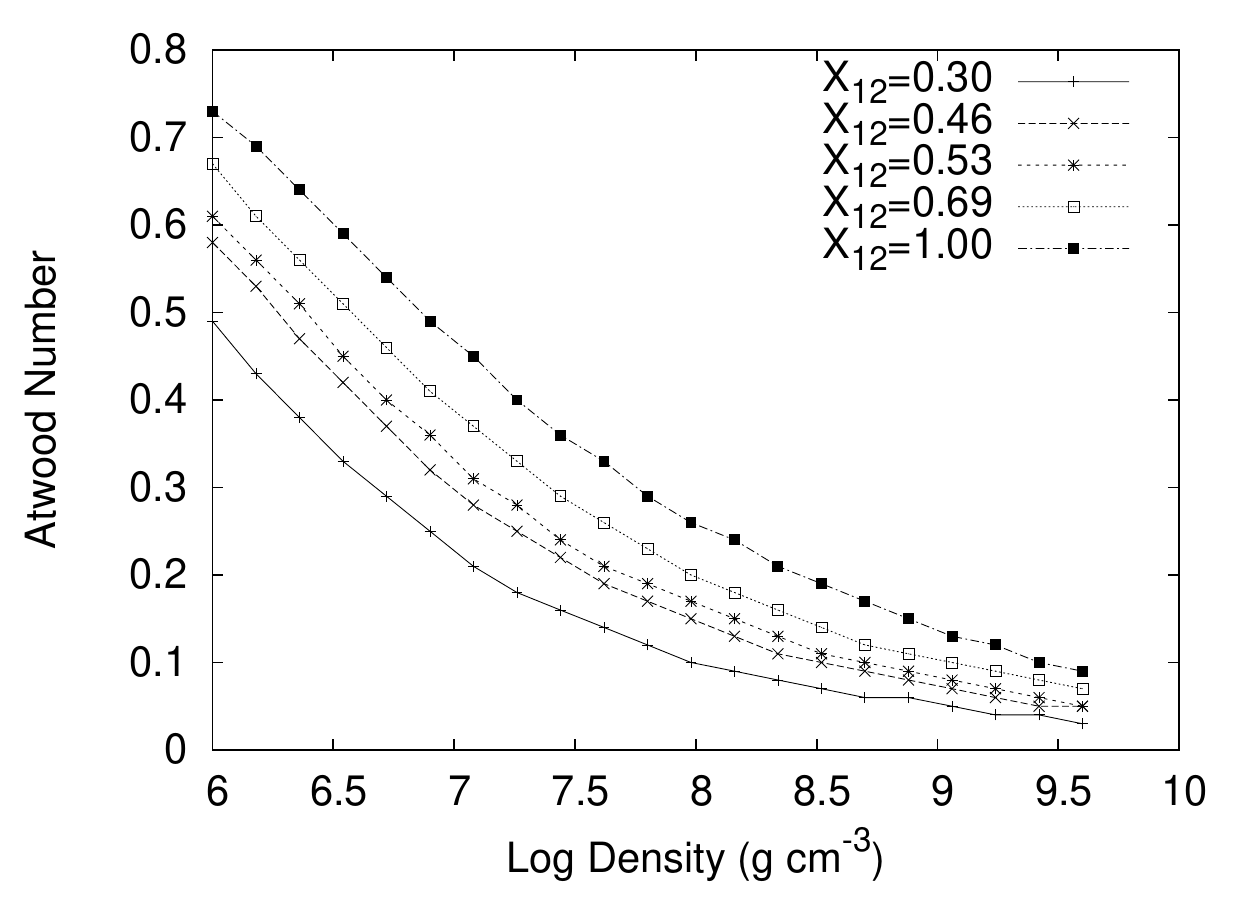}
%\plotone{figure1.pdf}
%\plotone{atwoodtab.pdf}
\caption{\label{fig:atwood}The Atwood number is shown as a function of log 
density for several carbon mass fractions.  This displays the sufficient 
smoothness of the Atwood number as a function of log density and carbon mass 
fraction such that our method of uncertainty estimation is valid.
}
\end{figure}

In order to test the accuracy of this interpolation procedure, we estimate
the uncertainty for the Atwood number throughout the table. Our method of
uncertainty estimation solves for the second-order term in a polynomial
interpolation of the table in each direction.  Because the Atwood number is 
a smooth and slowly varying function of $\XC$ and $\logten{\rho}$ as shown
in \figref{fig:atwood}, we can assume that higher order terms in the Taylor 
expansion are negligible.  In this case, the second-order term serves as a 
correction to the linear interpolation in a particular direction.  We use 
this correction to define the uncertainty estimate
\begin{equation}\label{errest}
R^A_\xi (\XC,\logten{\rho}) = - \left( \frac{ A^{\rm 2nd}_\xi - 
                                A^{\rm 1st} }{ A^{\rm 1st} } \right) \%
                                {\rm ,}
\end{equation}
where the Atwood number, $A = A(\XC,\logten{\rho})$, is calculated
using a 1st and 2nd order interpolation.  The second-order interpolation,
$A^{\rm 2nd}_\xi$, is performed in only one variable-dimension, 
$\xi = \{\XC,\logten{\rho}\}$. By subtracting the first-order from the 
second-order interpolation, we are left with a term proportional to 
the second partial derivative with respect to $\xi$. This allows us to assess 
the curvature of our table in each direction and whether we have enough points 
in our table in each dimension such that linearly interpolating the data 
provides an accurate estimate of the Atwood number. For the Atwood table, we 
are able to calculate an exact value to compare against the interpolated 
estimate.  However, for the flame speed table discussed later, we are not able
to calculate an exact value to test against.  We use the Atwood table to verify
our method of uncertainty estimation used later to analyze our flame speed table.  The 
comparison of the uncertainty estimate, $R^A_\xi$, to the actual error, 
$\sigma$, in \tabref{atwood_err} verifies our method of uncertainty 
estimation. 

The results of performing these tests at the grid points and midpoints of the 
Atwood table are shown in \tabref{atwood_err}. We show the minimum, maximum,
absolute average, and average uncertainty estimate, $R_{\rm max}, 
R_{\rm min}, \langle |R|\rangle, \langle |R| \rangle$, in each direction of the Atwood table. In 
columns 1 and 2, we evaluate the uncertainty estimate at a grid-edge (a
midpoint along $\xi$ holding the other variable at tabulated values).  In 
column 3, we evaluate the uncertainty estimate at a cell-centered point 
(where both variables are at midpoints).  In this case, we sum the uncertainty
estimates in each direction, $R^A_{\rm Sum} = R^A_{X12} + R^A_{\log{\rho}}$.  
We do not perform a bi-quadratic interpolation. While the cross-term is 
potentially important in this case, we do not include it in the uncertainty
estimate of a cell-centered point due to the dependence on the order of
interpolation in multi-dimensional polynomial interpolations.
In columns 4 and 5, we calculate the actual error, $\sigma$, for any $\XC$ 
and $\XC=0.5$, respectively. In this study, we use models with $\XC = 0.5$.
The error is calculated at all midpoints and grid points (the same values 
used for the uncertainty estimates).  Due to the way we initialize the
Atwood table, the test points are equivalent to forty-two equidistant 
log-densities between 6.0 and 9.6 and twenty equidistant carbon mass fractions
between 0.3 and 1.0 with $\XNe = 0.01$.

Because the table has positive curvature in $\XC$ and negative curvature in 
$\logten{\rho}$, we expect to find the maximum magnitude in the uncertainty 
estimate on a grid-edge as opposed to a cell-centered point. Our method of 
uncertainty estimation provides a maximum expected uncertainty of -10.99\% which 
is verified by the true maximum error of -9.99\%. Both of these values were 
evaluated at $\XC = 0.339$ and $\logten{\rho} = 9.422$. This shows that our
method of uncertainty estimation works for the Atwood table and can be 
applied to the flame speed table that will be discussed next. For the 
purposes of this study, the relevant maximum error is 
$\sigma^{\rm max}_{C12=0.5} = 7.6\%$ obtained from comparing the interpolation 
against calculated Atwood numbers at midpoint log-densities all at $\XC = 0.5$.

\begin{table}
\caption{\label{atwood_err}Error estimates for tabulating and linearly 
interpolating the Atwood number}
\begin{center}
\begin{tabular}{l||c|c|c|c|c}
$\xi$ & $\XC$ (\%) & $\logten{\rho}$ (\%) & $\XC + \logten{\rho}$ (\%) & 
$\sigma$ (\%) &  $\sigma_{C12=0.5}$ (\%) \\
\hline
$R_{\rm min}$ & -0.98 &  3.30 &  0.00 &  0.03 & -0.15 \\
$R_{\rm max}$ &-10.99 &  6.88 & -5.19 & -9.99 & -7.60 \\
$\langle |R| \rangle$     &  4.82 &  5.52 &  1.93 &  4.11 &  3.69 \\
$\langle R \rangle$       & -4.82 &  5.52 &  0.82 &  0.95 & -3.10 \\
\hline
\end{tabular}
\end{center}
\end{table}

\subsubsection{Composition Dependence of the Laminar Flame Speed}

\begin{figure}[t]
\includegraphics[width=\columnwidth]{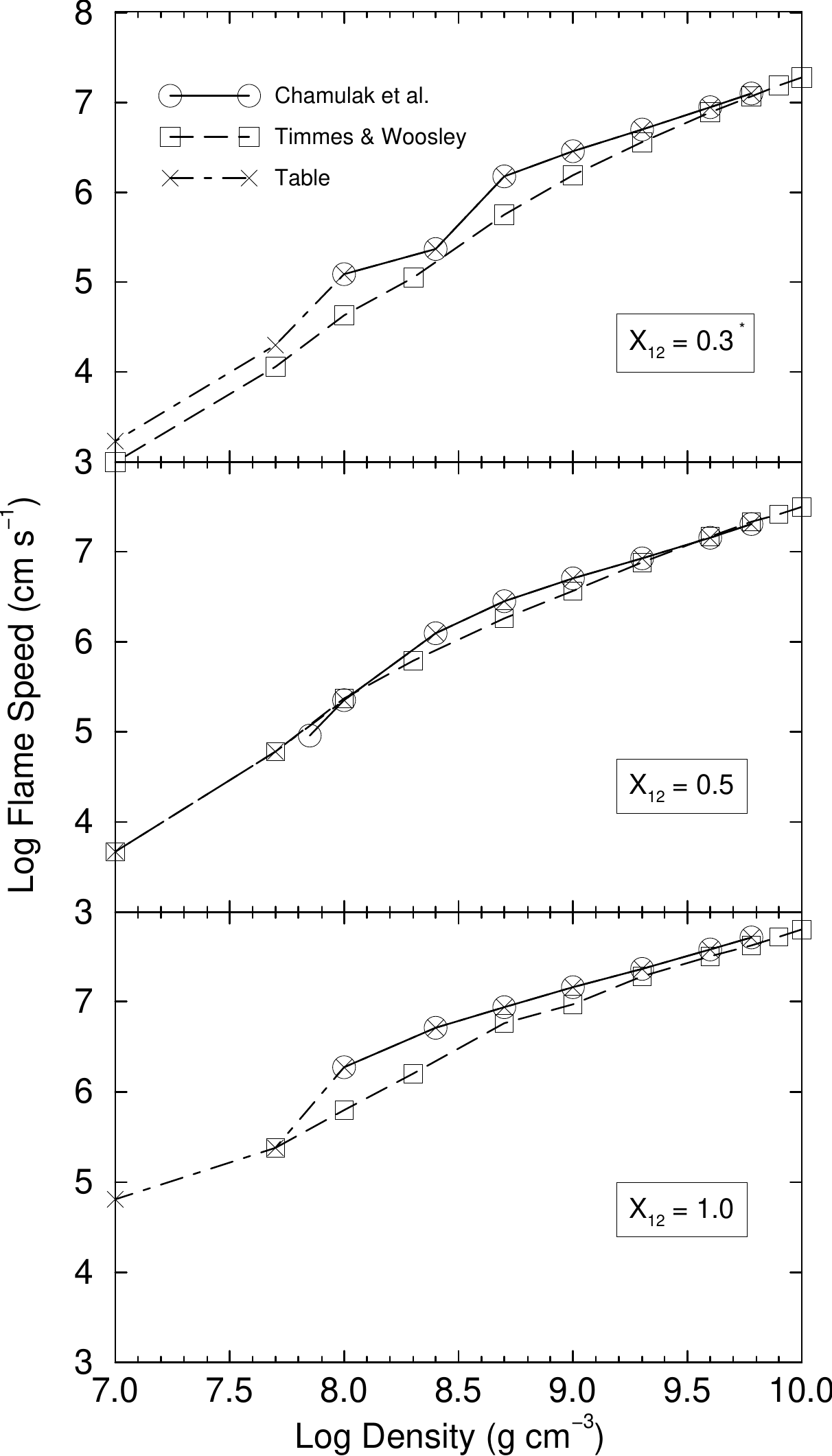}
%\plotone{figure2.pdf}
%\plotone{flamespeed3panel02.pdf}
\caption{\label{fig:fspd}Shown are laminar flame speeds as a function of log 
density from \citet{ChamBrowTimm07}, \citet{timmes_1992_aa}, and our 
resulting flame speeds from merging these datasets all at zero $^{22}$Ne mass
fraction.  Each panel compares these datasets at different carbon mass 
fractions, $X_{12}$. $^*$(Note that \citet{timmes_1992_aa} performed their 
study at $X_{12} = 0.2$. We expect slightly higher laminar flame speeds 
at $X_{12} = 0.3$.)
}
\end{figure}

For the laminar flame speed, we give preference to values calculated by 
\citet{ChamBrowTimm07} using a 430-nuclide reaction network for a variety
of initial carbon and neon mass fractions and a range of densities. Similarly
to the Atwood numbers, at runtime the code performs a linear interpolation to
obtain the laminar flame speed from a table of previously calculated results.
The method used by \citet{ChamBrowTimm07} is not well suited to solve for
the laminar flame speed at low density; therefore, we use the results from
\citet{timmes_1992_aa} to supplement this table.  Because the flame speeds
for each case were calculated using different initial carbon mass fractions,
we merged the data by linearly interpolating \citet{timmes_1992_aa} values
onto the \citet{ChamBrowTimm07} grid. The results of this merger are more
clearly shown in \figref{fig:fspd}.

Some discussion of the accuracy of this method of obtaining flame speeds at
low densities is warranted. \citet{timmes_1992_aa} do not track 
$^{22}$Ne dependence with their method of determining the laminar flame
speed. For identical points in parameter space, the two methods produce
laminar flame speeds that differ on average by $\sim 30\%$. This difference
is of the same order as the effect of adding $^{22}$Ne where we see a
$\sim 30-60\%$ speed-up depending on density.
For the models considered in this study, low densities
occur near the surface of the white dwarf star. In these regions, the input 
flame speed (\ref{fspd}) is dominated by the Rayleigh-Taylor driven 
turbulence such that $S_{\rm min} > S_{\rm lam}$. In fact, this transition 
occurs at $\rho \approx 2.5~\times~10^8$~g~cm$^{-3}$ in the initial star.
Therefore, this tabular 
method of estimating the laminar flame speed is sufficient at low densities 
for this study.

The tri-linear interpolation occurs within the three-dimensional parameter 
space of carbon and neon mass fraction and log-density to calculate the 
laminar flame speed. We cannot easily evaluate the uncertainty in our 
interpolation method by comparing with a direct calculation of the 
laminar flame speed due to the computational cost. Therefore, we apply our 
method of uncertainty estimation discussed for the Atwood number to the 
laminar flame speed table, Eq.~(\ref{errest}). We calculate an uncertainty estimate,
$R^S_\xi$, for our method of interpolation at the quarter-points and 
grid-points for each parameter in our table. We limit our analysis to densities
above $2.5~\times~10^8$~g~cm$^{-3}$. The laminar flame speed becomes 
unimportant below $2.5~\times~10^8$~g~cm$^{-3}$ because the 
buoyancy-compensated flame speed takes over in \eqref{fspd} at roughly this
density. The results of these calculations are given in \tabref{fspd_err} and 
\tabref{fspd_err_fixedcarbon}.  \tabref{fspd_err_fixedcarbon} shows results for 
$\XC = 0.5$, which is more relevant for this study, while \tabref{fspd_err} 
shows the behavior of our flame speed table in general. As was the case for 
the Atwood number, the uncertainty estimate along a single variable dimension, 
$R^S_\xi$, is evaluated along a grid-edge (at the quarter-points of $\xi$ 
holding the other variables at tabulated values). This is the case for 
columns 1-3. In column 4, we evaluate $\XC-\logten{\rho}$ cell-centered points 
(at the quarter-points of $\XC$ and $\logten{\rho}$ holding $\XNe$ at tabulated 
values). To evaluate the uncertainty estimate on cell-centered points, we sum 
the individual uncertainty estimates in each direction. In column 5, we 
calculate the uncertainty estimate at the quarter-points in all directions such 
that $\XC$, $\XNe$, and $\logten{\rho}$ are not at tabulated values.

For the flame speed table, both the $\XC$ and $\logten{\rho}$ variable 
dimensions have negative curvature, while $\XNe$ has mostly positive curvature.
Because the uncertainties along each grid-edge are comparable, we expect the 
maximum uncertainty in our table to occur either along a grid-edge or on a 
$\XC-\logten{\rho}$ cell-centered point. We do not expect a maximum uncertainty 
estimate involving $\XNe$ and any of the other two variables due to opposing 
curvatures.

\begin{table}
\caption{\label{fspd_err}Error estimates for tabulating and linearly 
interpolating the laminar flame speed (above $\logten{\rho} = 8.4$)}
\begin{center}
\begin{tabular}{l||c|c|c|c|c}
$\xi$ & $\XC$ (\%) & $\XNe$ (\%) & $\logten{\rho}$ (\%) & 
$\XC + \logten{\rho}$ (\%) & $\XC + \XNe + \logten{\rho}$ (\%) \\
\hline
$R_{\rm min}$ &   0.00 &  0.00 &   0.24 &  0.45 &   0.39 \\
$R_{\rm max}$ &   8.74 &-10.96 &   5.81 &  8.63 &   8.27 \\
$\langle|R|\rangle$ &   1.14 &  0.25 &   1.76 &  2.73 &   2.68 \\
$\langle R \rangle$       &   1.01 & -0.22 &   1.76 &  2.73 &   2.68 \\
\hline
\end{tabular}
\end{center}
\end{table}

\begin{table}
\caption{\label{fspd_err_fixedcarbon}Error estimates for tabulating and
linearly interpolating the laminar flame speed (above $\logten{\rho} = 8.4$ 
and $\XC = 0.5$)}
\begin{center}
\begin{tabular}{l||c|c|c}
$\xi$ & $\XNe$ (\%) & $\logten{\rho}$ (\%) & $\XNe + \logten{\rho}$ (\%) \\
\hline
$R_{\rm min}$ &  0.00 &  0.70 &  0.72 \\
$R_{\rm max}$ &  0.62 &  3.35 &  3.34 \\
$\langle |R| \rangle$     &  0.06 &  1.82 &  1.83 \\
$\langle R \rangle$       &  0.06 &  1.82 &  1.83 \\
\hline
\end{tabular}
\end{center}
\end{table}

We determined that the laminar flame speed table has sufficient 
resolution at densities above $\logten{\rho} = 8.4$ with the magnitude of 
estimated uncertainties $\lesssim 10\%$ as shown in \tabref{fspd_err}. The 
estimated uncertainties relevant for this study at $\XC = 0.5$ are 
$\lesssim 3\%$ as shown in \tabref{fspd_err_fixedcarbon}.

\subsection{Mesh Refinement}
\label{sec:refinement}

The goals of the design of our refinement scheme are to be as simple as
possible while both capturing interesting or physically important features
and doing so with good efficiency.  These three goals are somewhat at odds,
and therefore provide a wide latitude for choosing refinement prescriptions.
We proceed by defining three regions of the physical domain:
\begin{enumerate}
\item fluff: regions with $\rho < \rho_{\rm fluff}$
\item star: non-fluff regions, $\rho > \rho_{\rm fluff}$
\item energy generation:\\ regions with $\epsilon_{\rm nuc} >
\epsilon_{\rm eg}$ or $\dot\phi_{\rm RD}>\dot\phi_{\rm RD, eg}$
\end{enumerate}
where $\phi_{\rm RD}$ is the progress variable in the reaction-diffusion
model of the flame front.  These are indicated with a "$f$", "$*$" or "eg"
subscript respectively.
We assign to each of these a consecutively increasing maximum refinement
level.  For simplicity, we will here use the minimum cell size rather than
the refinement level, $\Delta_f>\Delta_*>\Delta_{\rm eg}$.

A ``fluff'' region outside the star is necessary because the hydrodynamics
method in FLASH has no explicit mechanism for treating empty (zero-density)
cells or free surfaces.  To ameliorate this, would-be empty cells are filled
with material of extremely low density which will not effect the dynamics of
the more dense material of interest in the simulation
 \citep[see also][]{zingale_2002_aa}.  Here we set it to
$10^{-3}$~g~cm$^{-3}$.
Because this material is of no physical interest and has negligible
contribution to the dynamics of other material, $\Delta_f$ is taken as large
as possible, generally being the total domain size divided by the block size.
Recall that the smallest independently refinable unit in PARAMESH is a
"block", which in all our simulations is a $16\times16$ cell region.

The finest resolution $\Delta_{\rm eg}$ will be treated as the resolution of
the simulation, though in fact before the burning spreads, most of the star
is only refinable to $\Delta_*$.  Some care must be taken in defining the
thresholds such that regions near the thresholds do not cycle between
refinement levels.  Cycling is avoided by checking both parent and child
blocks (i.e. blocks at the finest and next to finest refinement levels)
reconciling the results and using some hysteresis in defining the boundaries.
For the simulations presented here $\rho_{\rm fluff}=10^3$~g~cm$^{-3}$
throughout the simulation.  If a simulation is run far enough into expansion,
this restriction would need to be relaxed to lower density so that stellar
material remains refined as it expands.
The energy generating thresholds are placed at
$\epsilon_{\rm eg}=10^{18}$~erg~g$^{-1}$~s$^{-1}$ and $\dot\phi_{\rm
eg}=0.02$~s$^{-1}$ for the deflagration and detonation phases, but are relaxed
to $10^{19}$~erg~g~s$^{-1}$ and infinity, respectively, for calculations of the
expansion phase.

Beyond the definition of these regions of maximum refinement, refinement is
triggered by gradients in the physical quantities.  Here we use the density
and the first progress variable in the burning model, $\phi_{fa}$, to trigger
refinement.  This choice guarantees that all burning fronts are refined to the
degree allowable.  In order to detect gradients, we utilize the built-in
tests included with PARAMESH and FLASH.  These measure a ratio between the
second and first derivatives of the fields being checked.  The reader is
referred to the source code for the exact generalization to multiple
dimensions.  A threshold for triggering refinement is then defined for each
quantity checked, the default being 0.8.  In order for the initial star to be
fully refined, it was necessary to drop this threshold to 0.1 for the test of
the density field.  With this threshold, de-refinement set in during
expansion in one-dimensional simulations, but not appreciably so in two
dimensions.

\begin{figure}
\includegraphics[width=\columnwidth]{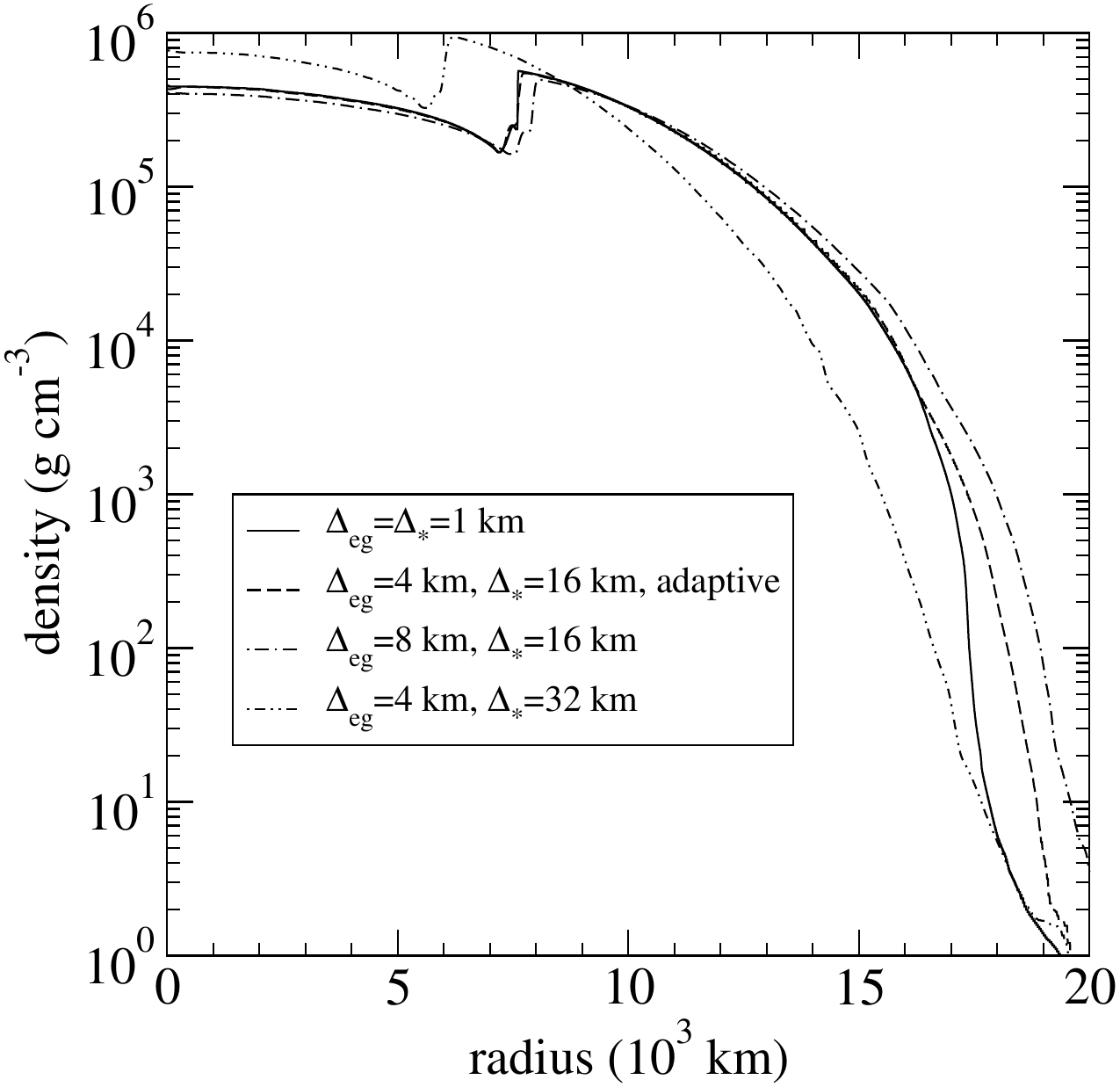}
%\plotone{figure3.pdf}
%\plotone{convergence_dens.pdf}
\caption{\label{fig:1dconvergence}
Density distribution of outgoing ejecta in radius for one-dimensional test of resolution.
Convergence is obtained when energy-generating regions are refined to 4~km
and the rest of the star is initially refined to 16~km (dashed line).  The
reference case in which the entire star is uniformly refined to 1~km is shown
by the solid line, while cases in which each of the refinement regions are
relaxed by one level from the coarsest converged values are shown by the
dot-dashed and dot-dot-dashed lines.  }
\end{figure}

In order to verify that sufficient resolution was being utilized and that the
adaptive refinement was not adversely affecting the accuracy of the
solution, a one-dimensional convergence test was performed.  Convergence is
expected in one dimension because instabilities are suppressed.  The absence of the
instabilities that accelerate the burning in a multidimensional simulation
necessitates a significant artificial enhancement of the burning rate in
order to unbind the star.  Because this is only a numerical test, we simply
increased the $m$ parameter (Equation [\ref{fspd_min}]) until an explosion was obtained.  Then the
combination $m\Delta$, where $\Delta$ is the finest resolution, generally
$\Delta_{\rm eg}$, is held fixed as the resolution is varied.  We used
$m\Delta=32$~km.  We are not interested in the order of convergence here,
only that reasonable convergence is obtained. Therefore, we simply compare
the density distribution of the outgoing ejecta to a uniformly refined case
with additional levels of refinement, $\Delta_{\rm eg}=\Delta_*=1$~km.  This
is the solid line in Figure~\ref{fig:1dconvergence}, which shows density
distributions for simulations at various resolutions at a time of 5.6 seconds
after ignition.  The fluff was not refined in any of these simulations,
but a test was performed with the entire domain refined to confirm that this
had no effect in the density range shown here.  The same initial WD was used
in all cases, which was created on an 8~km grid and mapped onto the new grid
by averaging density and temperature without a reconstruction (or
equivalently a piecewise constant one).

We found that a resolution of $\Delta_{\rm eg}=4$~km and $\Delta_*=16$~km was
necessary to satisfactorily match the reference solution.  Also it was
necessary for the initial star to be fully refined at $\Delta_*$, and the
trigger threshold for the density gradient test was adjusted to achieve this
as described above.  Subsequent de-refinement as the star expanded did not
appear to have any adverse effect on the outcome of the simulation.  The
de-refinement threshold was set to 0.0375.  The case with these parameters
matches the reference case extremely well down to a density of about $3\times
10^3$~g~cm$^{-3}$.  Most likely the initial resolution is insufficient to
resolve the hydrostatic equilibrium in these low density layers.  For
comparison, we show in Figure~\ref{fig:1dconvergence} the cases with a factor
of 2 too little resolution separately for the energy generation region and
the star.  We find the outcome is most sensitive to the refinement level of the star.
If this is too low, the hydrodynamics of the stellar expansion are not
properly captured.

This convergence study also provides an important check on the model of
nuclear energy release, particularly demonstrating that sub-stepping between
hydrodynamic steps does not appear necessary for such a simple reaction
model, as it likely would for even a highly reduced nuclear network.  It is
possible that lower resolutions of the energy generation region could be made
viable by introducing a sub-stepping mechanism.

%%%%%%%%%%%%%%%%%%%%%%%%%%%%%%%%%%%%%%%%%%%%%%%%%%%%%%%%%%%%%%%%%%
\section{Framework for Evaluating Systematic Dependencies}
\label{sec:framework}

As a starting point for our study of systematic effects, we must consider a
model of the supernova explosion that reproduces many observed
characteristics.  In light of its success in one-dimensional work
\citep[e.g.][]{HoefKhok96},
we take the deflagration-detonation transition as this starting point.  The
defining feature of this scenario is that the flame front will undergo a
transition to detonation when turbulent mixing on the scale of the flame
front becomes more vigorous compared to the propagation speed of the flame
front \citep{NiemWoos97,KhokOranWhee97}.
One-dimensional work such as \citet{HoefKhok96} utilized the density at
which the DDT occurs, $\rho_{\rm DDT}$, as a non-unique parameter.  Variation
of $\rho_{\rm DDT}$ then led to the observed variety of SN~Ia outcomes.
Instead of this, we make the assumption that the conditions that lead to the
DDT, while dependent on local properties (e.g., composition and turbulence
strength) are otherwise unique, though not currently known with precision.
Variety in the outcome for the same initial WD relies on the ignition
configuration being non-unique due to the turbulence present in the
convective core at the time of deflagration ignition.  In this section we
construct the essential component of our framework for evaluating systematic
dependencies: a theoretical population of SNe~Ia obtained by sampling a
defined ensemble of ignition conditions.  Some basic characteristics of this
population are discussed.  The next section, \ref{sec:2dddt}, describes a
typical outcome in detail along with additional technical details of the
implementation of the explosion.

The ability of isolated plumes to rise to low densities from a central
ignition before the star expands significantly, as demonstrated
by a number of studies
\citep{calder.ea:_offset_ignition_1,PlewCaldLamb04,calder.ea:_offset_ignition_2,LivnAsidHoef05,Plew07,Townetal07,RoepWoosHill07,Jordetal08},
presents a significant challenge to producing a realistic explosion with a
DDT.  If the transition to
detonation takes place as early as these simulations imply,
only a very thin layer of Si-group elements, the hallmark of a Ia
spectrum, are produced.  However,
\citet{RoepHilletal06} found that a multi-spot deflagration ignition led to
a much more symmetric deflagration.  Building upon this, somewhat like other
authors \citep{RoepNiem07,BravGarc08}, we here are able to
obtain realistic DDT explosions by considering deflagration ignition
conditions chosen to have no low-order power.

\subsection{Constructing a Theoretical Population of SN~Ia}
\label{sec:ignitioncondition}
\label{sec:systematics}

In order to evaluate systematic dependencies in the SN~Ia population, we
require a theoretical sample of supernovae that mimic the properties of
supernovae as a population.  SN~Ia possess an intrinsically large scatter in
the $^{56}$Ni mass synthesized during the explosion, ranging typically
between about 0.3 and 0.8$M_\odot$.  (See \citealt{howelletal+09} for one example
sample distribution.)  With the DDT scenario, we have found that this degree
of scatter can be obtained by introducing a certain degree of randomness into
the initial condition of the deflagration phase.  We now motivate and
describe this initial condition and then describe the samples
that it produces.

Our initial condition is motivated as a possible abstraction from prospective
three-dimensional studies of the very early deflagration phase.  During the first
$\sim0.1$~s of the deflagration, the flame surface will be heavily
influenced by the pre-existing convection field in the core.  In order to
develop a randomized sample of a variety of ignition conditions,
we choose to parameterize the possible outcomes of this early evolution
as harmonic structure in a flame surface when it reaches a modest distance
from the core.  We place the
position of the initial flame surface before perturbation at a radius of
$r_0=150$~km.  To this base radial position, perturbations with definite
harmonic content are
added in the form of spherical harmonics.  In the case of axisymmetry, only
$m=0$ spherical harmonics can be included, reducing to Legendre polynomials.
However, this technique should extend in a straightforward way to three
dimensions.  The initial position of the flame surface is defined to be
\begin{equation}
\label{eq:flaminit}
r(\theta) = 150{\ \rm km} +
30{\ \rm km}\sum_{\ell=\ell_{\rm min}}^{\ell_{\rm max}} A_\ell
Y_{\ell}^{0}(\theta)\ .
\end{equation}
The normalization convention of \cite{Jack99} is used.   The amplitudes
$A_\ell$ are chosen from a normal distribution.  The maximum harmonic content
$\ell_{\rm max}$ is chosen so that the perturbations in the flame surface are
resolved.  For the 4~km resolution simulation performed here, we choose
$\ell_{\rm max}=16$.  As demonstrated by a number of studies of localized,
off-center ignition in the absence of a convection field
\citep{calder.ea:_offset_ignition_1,PlewCaldLamb04,calder.ea:_offset_ignition_2,Plew07,RoepWoosHill07,Townetal07,Jordetal08},
low-order perturbations on such a flame surface would lead to an early DDT
with too little expansion of the WD to give realistic yields.  To prevent
this effect, low order modes are left out by choice of $\ell_{\rm min}$.
Here we have used $\ell_{\rm min}=12$, which appears to give a reasonable
sample (see below).  An extensive study of the sample that arises from
different $\ell_{\rm min}$ values was not undertaken, so the sensitivity of
the sample to this choice is uncertain.  One additional restriction was
imposed due to the axisymmetry.  The perturbation, the second additive term
in equation (\ref{eq:flaminit}), is restricted to be negative on the symmetry
axis, $\theta = 0$, $\pi$.  This suppresses the formation of slender "jets"
of burned material, which are likely artifacts of the axis singularity.  Such
jets are much smaller than the rising bubbles observed by \cite{Townetal07}
and similar simulations, and bear more resemblance to the "tails" seen in
those cases.

Within the framework defined by equation (\ref{eq:flaminit}), besides
$\ell_{\rm min}$ and $\ell_{\rm max}$, it in necessary to specify the
implementation of the random choices of the $A_{\ell}$.\footnote{Our
implementation of the process described here is available from
{\tt http://variable.as.arizona.edu/code}.}
We will refer to a
set of these $A_\ell$ as a realization of the initial condition.
Rather than choosing completely unrelated random seed for each realization of
the initial condition,  a more well-defined and reproducible sample is
obtained by drawing random numbers for each consecutive realization from a
single random number stream.  Thus the entire sample is represented by the
initial seed of the random number stream, along with algorithmic details.
Also this gives a definite ordering to the realizations, arising from the
order of the random number stream.  We do not require a large number of
random numbers, and therefore opt for a simple linear congruential generator
(LCG) pseudo-random number generator with a 31-bit seed/output value.
The ending seed from one realization is simply used as the starting seed for
the next.  Candidate realizations that do not satisfy the property of having
a negative perturbation on the symmetry axis are dropped and another is
generated.  These dropped candidates are not included in the numbering of the
realizations used in the next section.  We use the LCG discussed in section
16.1.3 of \cite{NewmBark99}.  The initial seed for the set of realizations
presented here was obtained from the standard Linux kernel random number
source (/dev/random) and is 1866936915.

These sets of initial conditions provide a framework within which we may study
a wide variety of possible systematic effects.  This includes both physical
systematics such as those explored in this study, or those arising from
physical uncertainties in the numerical model.  As study of the central
ignition mechanism for SNe~Ia advances, with improved flame models and DDT
conditions, for example, it may be necessary to adjust the overall parameters
of this framework, notably $\ell_{\rm min}$, in order to maintain a realistic
sample.  Also, as discussed previously, these choices can be compared to
simulations of the early deflagration phase in order to both inform future
choices and retrospectively understand the context of previously performed
studies.  One advantage of such a controlled initial condition is that it
provides a slightly stronger probe of systematic effects than would generally
be available observationally.  This arises from the fact that it is possible
to compare the same ignition sample rather than independent samples, though
the latter can also be performed if desirable.

\subsection{The Theoretical Population}
\label{sec:population}

The above development provides a clear definition of a population of ignition
conditions from which we may draw a sample.  Lacking further information
about the nature of the harmonic content of the initial condition, we will
assume cases drawn from this sample will have equal weight (likelihood).
The population of supernovae that results is a purely theoretical construct.
Any observed population will have a variety of progenitors that will have a
distribution of intrinsic properties (composition structure, accretion
history, etc).  Therefore, while this population of ignition conditions will
be very useful for studying systematic effects, caution should be exercised
when drawing conclusions about observational ramifications.  Future studies
will address a more observationally motivated sample.

In order to understand the diversity in our sample, and the qualitative
changes arising from changes in $^{22}$Ne fraction, the first 5 realizations
from the sample sequence were run through the end of the detonation phase.
The basic outcome properties of these ten cases are listed in table
\ref{tab:detsample}.  The DDT time, $t_{\rm DDT}$, is defined as the time at
which any part of the flame surface first passes a density of
$10^7$~g~cm$^{-3}$.  This is the time at which the first detonation front is
launched.  The mass at high density at the DDT time, in this case that above
a density of $2\times 10^7$~g~cm$^{-3}$, will be discussed more below.  The
NSE mass, $M_{\rm NSE}$, is defined as $\int \phi_{qn}\rho dV$ integrated
over the star.  The $^{56}$Ni mass is determined by using eq.
(\ref{eq:xni56}) below to estimate the local mass fraction from the $Y_e$ and
again integrating over the whole ejecta.
The Si-group and O-Si masses are defined as $\int (\phi_{aq}-\phi_{qn})\rho dV$ and
$\int (\phi_{fa}-\phi_{aq})\rho dV$ respectively and are discussed more in section
\ref{sec:2dddt}, where the total nuclear energy released
$\Delta E_{\rm rest~mass}$ is also defined.
All of these yields are determined
after the detonation phase is complete, and the gross
production of burning products is complete (see section \ref{sec:2dddt}).
The amplitudes of the additive components that
make up the initial conditions are given in Table \ref{tab:amplitudes},
though this does not bear out any particular pattern in the results under
discussion.

\begin{table*}
\caption{\label{tab:detsample}Outcomes for subset of theoretical sample}
\begin{center}
\begin{tabular}{c|ccccccc}
realization &
  $t_{\rm DDT}$ [s] &  $M(\rho_7>2,t_{\rm DDT})$ $[M_\odot]$ &
 $M_{\rm NSE}$ $[M_\odot]$ & $M_{^{56}\rm Ni}$ $[M_\odot]$ &
 $M_{\rm Si-group}$ $[M_\odot]$ & $M_{\rm O-Si}$ $[M_\odot]$ &
 energy released [$10^{51}$ erg]\\
\hline
 \multicolumn{8}{c}{$X_{^{22}\rm Ne}=0$}\\ 
\hline
1  &
   1.17 & 1.00 & 0.91 & 0.79 & 0.34 & 0.11 & 1.93\\
2  &
   1.19 & 0.96 & 0.90 & 0.78 & 0.34 & 0.11 & 1.92\\
3  &
   1.36 & 0.78 & 0.75 & 0.64 & 0.45 & 0.14 & 1.88\\
4  &
   1.36 & 0.77 & 0.66 & 0.55 & 0.53 & 0.15 & 1.85\\
5  &
   1.16 & 0.96 & 0.89 & 0.77 & 0.36 & 0.11 & 1.92\\
\hline
\multicolumn{8}{c}{$X_{^{22}\rm Ne}=0.02$} \\
\hline
1  &
   1.21 & 0.87 & 0.88 & 0.72 & 0.35 & 0.11 & 1.93\\
2  &
   1.15 & 0.97 & 0.84 & 0.68 & 0.38 & 0.13 & 1.91\\
3  &
   1.43 & 0.54 & 0.60 & 0.46 & 0.57 & 0.17 & 1.83\\
4  &
   1.29 & 0.81 & 0.73 & 0.58 & 0.48 & 0.14 & 1.86\\
5  &
   1.17 & 0.92 & 0.83 & 0.67 & 0.40 & 0.13 & 1.92\\
\hline

\end{tabular}
\end{center}
\end{table*}

\begin{table}
\caption{\label{tab:amplitudes} Initial condition amplitudes}
\begin{center}
\begin{tabular}{c|ccccc}
realization & $l=12$ & 13 & 14 & 15 & 16\\
\hline
1 & -1.20 & -1.16 & 0.49 &  0.21 & -1.16 \\
2 & -0.73 & 1.17 & 0.95  & -1.18 & -0.62 \\
3 & -0.94 & -0.64 & 0.69 & 0.14 & -0.61 \\
4 & -1.15 & -1.00 &  0.03 & 1.11 & 0.60 \\
5 & 0.70  & -1.69 & -1.17 & 0.29 & -1.05 \\
\hline
\end{tabular}
\end{center}
\end{table}

Even this small sample from the theoretical population produces a diverse set
of supernovae.  The estimated yield of $^{56}$Ni spans a range between 0.45
and 0.8$M_\odot$.  The average is in the upper portion of this range.  For
$X_{^{22}\rm Ne}=0$ the average is $0.70\pm0.05M_\odot$, with a sample
standard deviation ($\sigma$) of 0.11$M_\odot$.  For $X_{^{22}\rm Ne}=0.02$ 
the average
is $0.62\pm0.05M_\odot$, also with a sample standard deviation of
0.11$M_\odot$.  The decrease of the $^{56}$Ni yield with increasing $^{22}$Ne
content is also directly reflected in the fraction of the NSE material that
becomes $^{56}$Ni.  This fraction is 0.86 with $\sigma = 0.01$ for
$X_{^{22}\rm Ne}=0$, but is 0.80 with $\sigma = 0.02$ for $X_{^{22}\rm
Ne}=0.02$.  This is a direct result of the differences in initial neutron
fraction due to the presence of $^{22}$Ne.

\begin{figure}
\includegraphics[width=\columnwidth]{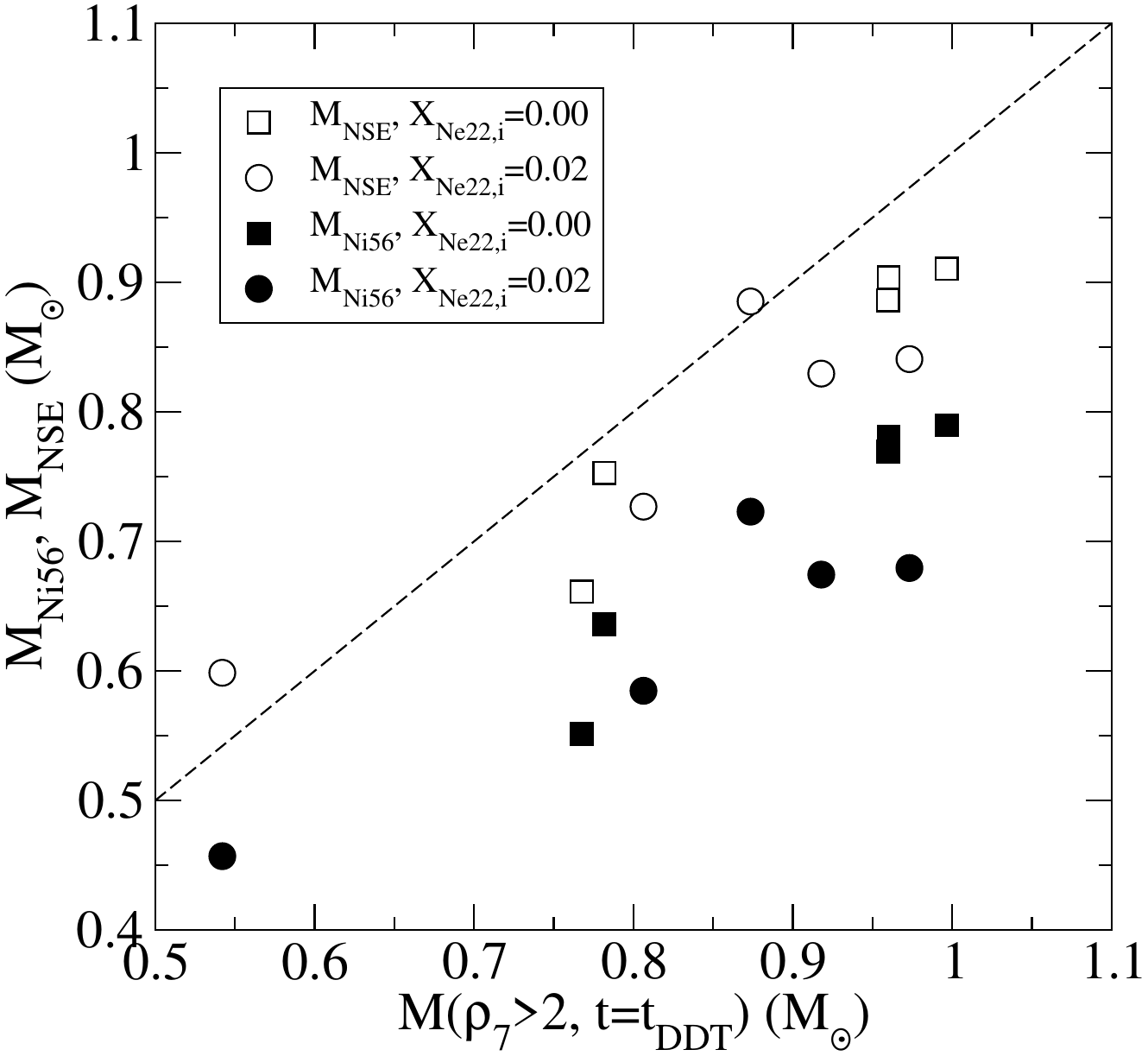}
%\plotone{figure4.pdf}
%\plotone{Mhi_Mni_Mnse.pdf}
\caption{\label{fig:Mhi_Mni_Mnse}
Mass of all Fe-group (NSE, open symbols) elements and estimated mass of
$^{56}$Ni (solid symbols) produced in the explosion plotted against the mass
of material above a density of $2\times 10^7$~g~cm$^{-3}$ when the first
detonation begins, an indicator of the degree of expansion of the star prior
to the DDT.  Two initial abundances of $^{22}$Ne are included,
$X_{^{22}\rm Ne}=0$ (squares) and $X_{^{22}\rm Ne}=0.02$ (circles).  Both of
these yields are well-correlated with the mass at high density at the DDT
transition, making the latter a useful intermediate indicator of the
explosion outcome.
}
\end{figure}

The main determining factor in the gross amounts of products synthesized in
the explosions is the degree of expansion when the detonation ignites.
This dependence has been discussed by previous authors
\citep[e.g.][]{RoepNiem07,Townetal07,Meaketal09}, but the details of such a relation will
differ among different delayed detonation scenarios.  It is useful to
characterize the degree of expansion by the mass at high density (above some
density threshold), which forms an indicator of how much material will be
processed to NSE.  A part of this will become $^{56}$Ni and determine the
overall brightness of the supernova.  Several density thresholds for defining
the mass at high density were considered. The mass above a density of
$2\times 10^7$~g~cm$^{-3}$, $M(\rho_7
>2)$, appears the most appropriate.  This conjecture is demonstrated 
by the open symbols
in figure \ref{fig:Mhi_Mni_Mnse}, which lie near a 1-1 relation (dashed
line).

\begin{figure*}
\centerline{ \includegraphics[width=0.88\textwidth]{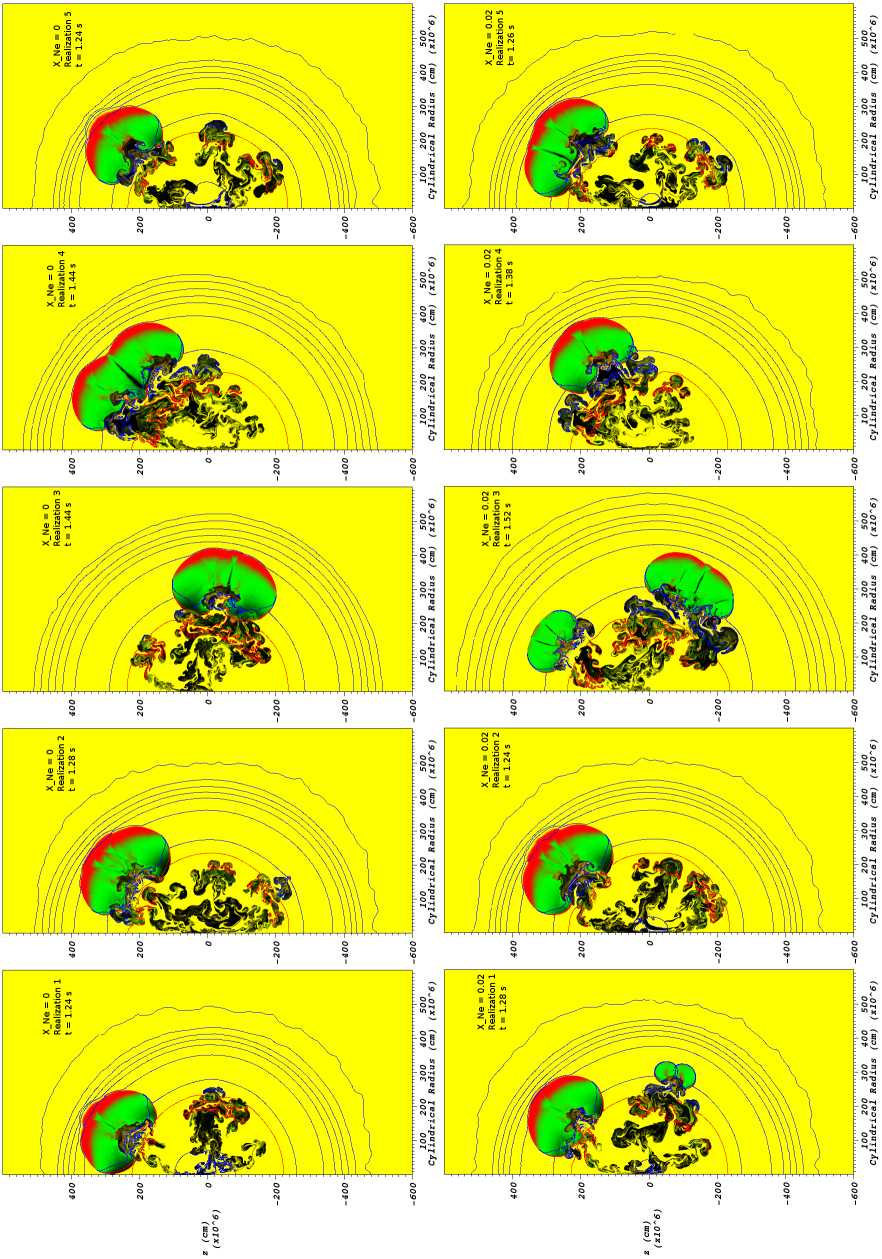}}
%\epsscale{0.95}
%\plotone{figure5.jpg}
%\plotone{ddt_compare.png}
\epsscale{1.0}
\caption{\label{fig:ddt_compare}
Comparison of burning products approximately $0.1$ seconds after first
detonation is launched for different realization of the initial flame surface
(columns) and for two abundances of $^{22}$Ne (rows).  Simulations are
performed in axisymmetry.
Fuel and burning products are indicated by color: unburned C, O, Ne ({\it
yellow}), O-Si ({\it red}), Si-group ({\it green}), Fe-group (NSE, {\it
black}).
Density in g~cm$^{-3}$ is indicated by contours logarithmically spaced at
integer powers of 10 starting from $10^1$~g~cm$^{-3}$ at the outside.  One
extra contour ({\it red}) is added at a density of $2\times
10^7$~g~cm$^{-3}$.
}
\end{figure*}

The variation in expansion at the DDT appears to arise from a competition
between the rise of the highest plume and the expansion of the star in
response to energy input.  Both these processes have inherently similar
timescales, the star's dynamical time.  Figure \ref{fig:ddt_compare} compares
the state of the star approximately 0.1 seconds after the launch of the first
detonation for the 5 realizations at two $^{22}$Ne fractions run through the
detonation phase.  The top row displays the $X_{^{22}\rm Ne}=0$ and the
bottom $X_{^{22}\rm Ne}=0.02$, with each column being a separate realization
of the ignition condition.  The coloring indicate the nucleosynthetic yield,
with black indicating NSE material, green Si-group material, and
red O-Si mixture.  It is immediately evident that the lowest $^{56}$Ni-yield
case, realization 3 for $X_{^{22}\rm Ne}=0.02$, is also that which is the
most expanded at the DDT transition.

The outcome of the deflagration appears to arise to some degree from the
morphology of the deflagration, and therefore presumably from (randomly
determined) characteristics of the ignition condition.  Cases for which the
highest plume is near the axis, realization 1, 2 and 5, expanded less before
the detonation began than more equatorial lead plumes as in realization 3.
Realization 4 appears to form an intermediate case, which in turns makes its
outcome the most sensitive to the inclusion of $^{22}$Ne among these trials.
This effect can be reasonably understood in terms of the interaction of the
Rayleigh-Taylor (R-T) instability with the axisymmetric geometry.  Plumes
near the equator are dynamically more like rings instead of columns,
therefore being more two-dimensional structures, and therefore are driven
somewhat more weakly by the R-T instability.  This effect may account for their
slower rise in these simulations, and the resultant longer delay before
detonation ignition.  Additionally, equatorial plumes will generally burn
material more quickly due to a larger integrated surface and volume in
axisymmetry, and therefore could also enhance the expansion of the star
directly.

This interaction with the geometry is important for extending these results
to three dimensions.  While the direct differential suppression of R-T will
no longer be important, the presence or absence of localized "spikes" in the
initial condition is likely to become the most important determining factor
in the competition between stellar expansion and plume rise, and therefore
the explosion yield.  Thus, assumed geometry is no longer important, but
geometric features of the initial condition serve a similar role.  This
also creates the possibility that the specific morphology imparted to the
flame surface by the convection field in the early deflagration phase is an
essential aspect of the outcome of the DDT scenario.  Filamentary or
sheet-like structures might act quite differently from lumps of burned
material once the strong R-T growth phase takes over.

%%%%%%%%%%%%%%%%%%%%%%%%%%%%%%%%%%%%%%%%%%%%%%%%%%%%%%%%%%%%%%%%%%
\section{A Deflagration Detonation Transition Supernova in two dimensions}
\label{sec:2dddt}

In this section we describe the outcome of a typical case from the ensemble
that will be used to study systematic effects,
the case labeled "realization 2" or "r2" for
$X_{\Ne}=0.02$.  This case is considered typical because the mass at high
density at the time of transition to detonation (see Section~\ref{sec:population})
and therefore the total mass of $^{56}$Ni synthesized, $0.68M_\odot$, is
similar to the most probable values for both the ensemble of section
\ref{sec:framework}, and, for $^{56}$Ni mass, the observed distribution
of SNe~Ia \citep[e.g.,][]{howelletal+09}.

\subsection{The Explosion}
The explosion can be roughly divided into 3 phases, the deflagration phase,
the detonation phase, and the expansion phase.  The deflagration phase lasts
for the first 1.2 to 1.4 seconds of the simulation, depending on the
particular ignition condition.  During this phase a subsonic burning front,
accelerated by turbulence and buoyancy, burns material in the inner portion
of the star and expands it appreciably.  As plumes of burning material rise
from the core, the deflagration front eventually reaches low enough densities
to transition to a detonation.  The first of these to initiate a detonation
will begin the detonation phase, which will burn the entire star within a few
tenths of a second.  Once all the material is burned, the star will continue
to expand, beginning the expansion phase.  The outgoing ejecta will
eventually reach a state of free expansion in which the radial
velocity of material is a linear function of radius and therefore the
density becomes a simple function of time.  We will now discuss each of these
stages in turn for a typical case from our ensemble, including some details
about how the simulation is accomplished.

\subsubsection{Deflagration}

\begin{figure*}
\epsscale{0.85}
\plotone{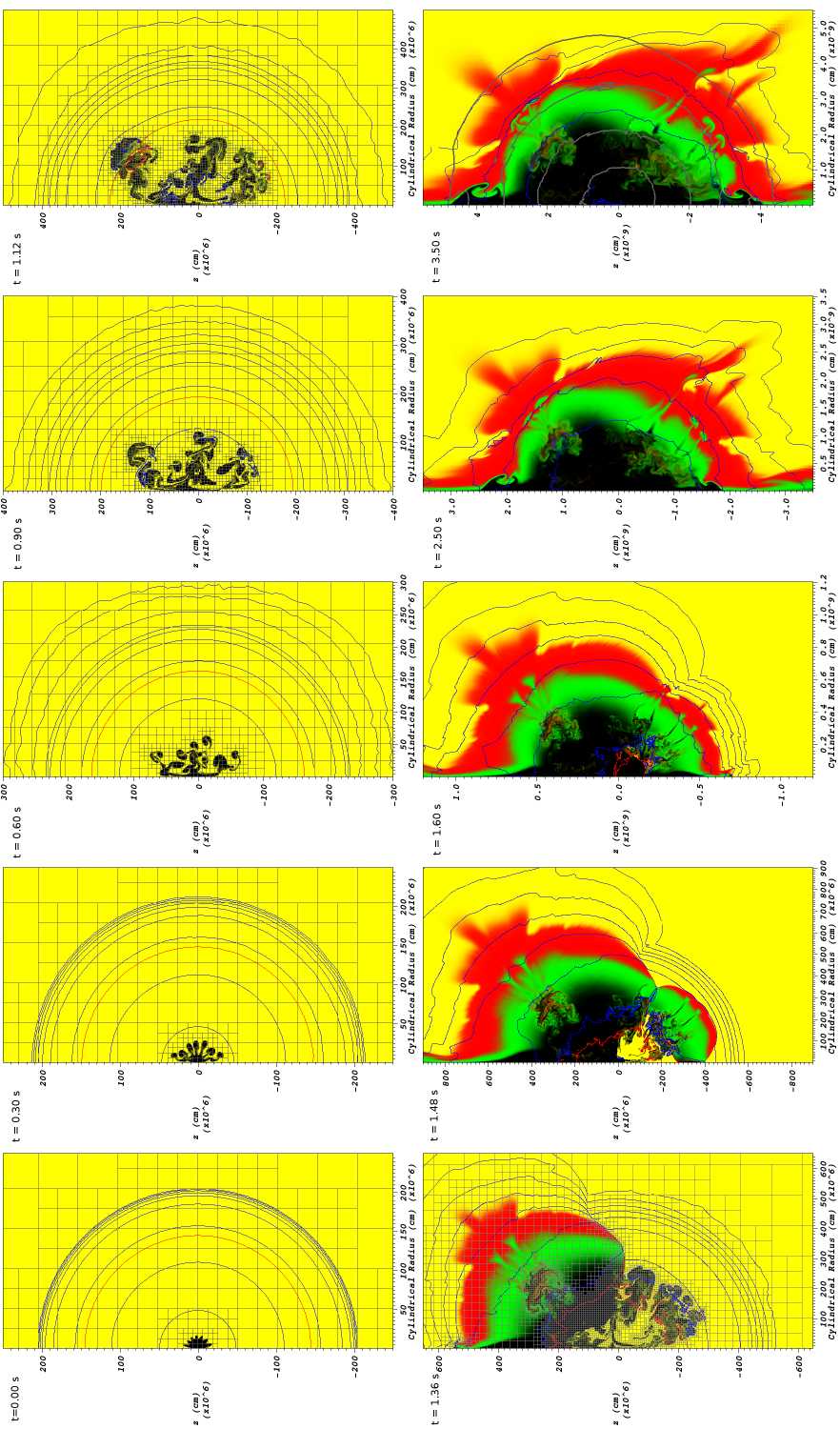}
%\plotone{r2_sequence.png}
\epsscale{1.0}
\caption{\label{fig:sequence}
Time evolution of star and burning products during the explosion and
establishment of the expansion.  This sequence is for realization 2 with
$X_{\Ne}=0.02$ from the sample shown in Figure~\ref{fig:ddt_compare}, which is a typical
case based on exploration of our sample in section~\ref{sec:population}.
By the end of this sequence ($t=3.5$~s) all
burning has ceased and the star is nearly in free expansion.  Fuel and
burning products are indicated by color: unburned C, O, Ne ({\it yellow}),
O-Si ({\it red}), Si-group ({\it green}), Fe-group (NSE, {\it black}).
Density in g~cm$^{-3}$ is indicated by logarithmically spaced at integer
powers of 10 starting from $10^1$~g~cm$^{-3}$ at the outside.  One extra
contour ({\it red}) is added at a density of $2\times 10^7$~g~cm$^{-3}$.  The
final panel also includes contours of radial velocity at 5, 10, 15, and
20~km~s$^{-1}$ ({\it thick gray}).  The adaptive mesh is indicated in the
early panels by outlines of blocks of $16\times 16$ cells, the smallest unit
of contiguous refinement.  The first detonation points are initiated at the
moment depicted in the 5th panel at $t=1.12$~s.
}
\end{figure*}
The initial burned region and the progression of the deflagration phase for
realization 2 is shown in the top row of Figure \ref{fig:sequence}.  The
burned material is colored black throughout this phase, because burning in the
flame at these densities always results in Fe-group (NSE) material within the
width of the artificial flame.  Black lines show density contours on a
logarithmic scale in integer powers of 10 beginning at 1 in the outer regions
and reaching to 9 in the first panel.  The initial WD has a central density
of $2.2\times 10^9$~g~cm$^3$.  The adaptive grid is shown by outline of the
"blocks." These are logical mesh regions of $16\times16$ cells that
represent the smallest region that can be independently refined to the next
level.  Note also that PARAMESH restricts neighboring refinement regions to
differ by at most a single refinement level (a factor of 2 in cell size).  As
discussed in section \ref{sec:refinement}, the background star is refined to
16~km resolution, while energy-generating regions are refined to 4~km.  The
"fluff" outside the star is not refined except as necessary to accommodate the
interior grid regions.

\subsubsection{Detonation and Unbinding}

In the simulations, the transition of the burning 
to detonation must be initiated artificially
because thermally activated burning is explicitly suppressed within the
RD front (the artificially broadened flame).  While section \ref{sec:burning}
describes a method for allowing thermal burning within the RD front for the
purpose of allowing the detonation to propagate into material that has been
"partially burned," the necessary estimate of temperature has so far proven
to be of limited utility beyond detonation propagation.  Spurious detonations
typically occur if the thermal burning proximity detection is not utilized.
Detonations are initiated by setting $\phi_{fa}$, the progress variable that
represents the consumption of $^{12}$C, to 1 in one or several neighboring
zones in one time step.  Note that this involves no addition of energy to
raise the temperature.  The increased pressure resulting from the energy release
following from the change in composition can then initiate a detonation that
propagates away from the ignition point.  It was found that for detonations at
the density used here ($10^7$~g~cm$^{-3}$), lighting a single $4\times 4$~km
cell does not always successfully launch a detonation.  The outcome of small
ignition points appeared to depend on the local flow characteristics, with
fast-rising plumes being one of the more commonly problematic regions.  This
problem could be ameliorated by simply increasing the size of the lighted
region.  We found good success using a region with a linear radius of 8~km,
resulting in the simultaneous ignition of a $5\times5$ cell region, still
small compared to the overall flow characteristics and scale for changes in
density.

For this study we have chosen to characterize the DDT point by a simple
density criteria, $\rho_{\rm DDT} = 10^7$~g~cm$^{-3}$.  Whenever the flame
front, as represented in the simulation by the RD front, reaches this
density, we light a detonation.  Because the top of the deflagration is
generally characterized by a few dominant plumes, the positioning of the
detonation ignition points is relatively unambiguous.  While $\rho_{\rm DDT}$
quantitatively defines our detonation point, it is further necessary specify
the method in which this density is used for placing the ignition point or
points.  In order to increase the chance of obtaining a robust detonation
ignition, the ignition point is placed slightly outside of the rising burned
region.  Points are ignited when the RD front is approximately 64 km below
the $10^7$~g~cm$^{-3}$ contour, at a point halfway between the RD front and
the contour.  Typically two points are ignited per plume, but sometimes three if
the plume is wide, as is the case for the first ignited plume in realization
2 shown in Figure \ref{fig:sequence}.  This method places the detonation
points at a density of between 1.05 and 1.1$\times 10^7$g~cm$^{-3}$.  A
density contour of $2\times 10^7$g~cm$^{-3}$ is shown in Figure
\ref{fig:sequence} in order to show that the ambiguity in detonation time and
place, introduced by the slight difference between $\rho_{\rm DDT}$ and the
actual point of detonation ignition, should be a fairly small factor in the
outcome of the overall explosion, assuming it is applied consistently.

It should be emphasized that this treatment of detonation ignition is only
a first, simplest option.  The correct placement of the detonation ignition
point is currently unknown, but at minimum a more realistic condition would
take into account the local flow characteristics in an estimate of the Gibson
scale and compare it to the flame width.  It is even possible that the
detonation ignition is qualitatively different.  For example, we have ignited
the detonation on the "top" of the rising plumes, although the most
pronounced mixing and tumble is occurring on the underside of plume caps.  It
may be that the detonation does not occur until this region reaches some
density threshold, as mixing on the plume tops is less vigorous.  This condition would
lead to a systematic delay of the detonation ignition from the methodology
implemented here.  We leave evaluation of various alternatives for ignition
of the detonation to future work.

Once the detonation is ignited, most of the remainder of the star is burned
in a few tenths of a second, between 1.12 s and 1.6 s in realization 2.  In
the sequence shown in Figure \ref{fig:sequence}, additional ignition points
launch from where plumes in the lower half-plane reach $\rho_{\rm DDT}$.  The
refinement tracks the detonation at lower densities, but much of the higher
density material remains refined longer because of the additional energy
release as NSE material expands and alpha particles are recaptured.  The
display of the grid has been eliminated after the sixth panel due to it
becoming too dense in this visualization.  Nearly all of the material in the
previously deflagrated region is burned, mostly to Fe-group material,
demonstrating the success of the method used to allow the detonation to
propagate into material within the RD front.  Some material in heavily mixed
regions is not fully burned, but given the low resolution achievable in the
simulations, it is unclear how realistic this is.  This incomplete
burning does not appear to be
a significant issue for gross yields from the explosion, but detailed
nucleosynthesis, which will be undertaken in future work, will need to
provide a better treatment of how deflagration ashes interact with the
detonation front.  This is discussed more with the final yields below.

\begin{figure}
%\epsscale{1.1}
\plotone{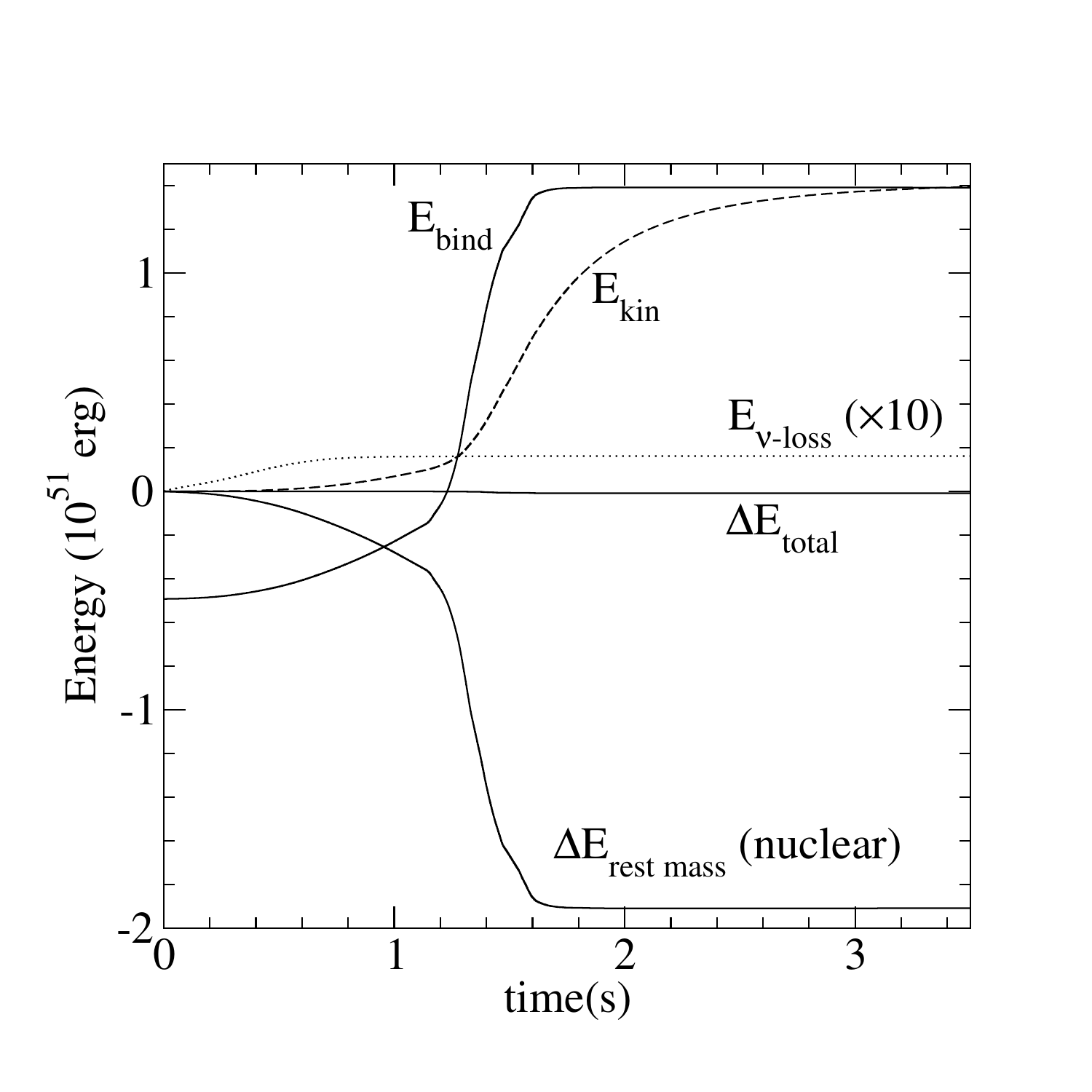}
%\plotone{energies_02_r2.pdf}
\caption{\label{fig:energy}
Dynamics of energy content during explosion.  Shown are the total
binding energy, $E_{\rm bind}$, including the gravitational binding energy,
the internal energy and the kinetic energy, the kinetic energy as a separate
component, $E_{\rm kin}$ ({\it dashed}), the accumulated energy lost in the
form of neutrinos, $E_{\rm \nu-loss}$ ({\it dotted}), the change in the total
rest mass energy of the star due to nuclear processes, $\Delta E_{\rm rest\
mass}$ and the change in the total of all these energetic components $\Delta
E_{\rm total}$.  The term $E_{\rm \nu-loss}$ has been multiplied by 10 for
display purposes.
}
\end{figure}
The energetic history of the explosion is shown in Figure \ref{fig:energy},
where we find that, as expected, most of the energy is deposited during the
detonation phase, and that the star is, in fact, bound up to this point.
Shown is the total binding energy, $E_{\rm bind}$, which includes the
gravitational binding energy, the internal energy, and the kinetic energy.
The exclusive source of energy is the nuclear energy release.  The source of
this energy is actually the change in rest mass energy of material due to
nuclear processes.  We quantify this by measuring the difference between the
rest mass of all the material on the grid and the same number of baryons of
symmetric matter ($Y_e= 0.5$) in a completely unbound state of free protons,
neutrons and electrons:
\begin{eqnarray}
\frac{e_{\rm rest\ mass}}{N_A} &=& 
Y_e(m_p+m_e)c^2 + (1-Y_e)m_n - \bar q\nonumber\\
&&\quad\mbox{}
-\left[0.5(m_p+m_e)+0.5m_n\right]\nonumber\\
&=&
 (Y_e-0.5)(m_p+m_e-m_n)c^2 -\bar q\ ,
\end{eqnarray}
where $e_{\rm rest\ mass}$ is the rest mass energy per $N_A$ baryons
(approximately 1 g of baryons), $m_p$, $m_n$, and $m_e$ are the rest masses
of the proton, neutron and electron respectively, and $\bar q$ is the average
nuclear binding energy as defined in section \ref{sec:burning}.  The total
rest mass energy is then $E_{\rm rest\ mass} = \int e_{\rm rest\ mass}\rho
dV$, where the integral is over the computational domain.  As discussed in
section \ref{sec:burning}, we are using the convention that density as
represented on the grid is actually the baryon density divided by $N_A$, and
is only approximately the density in g~cm$^{-3}$.  Due to the binding energy
of the unburned material, $E_{\rm rest\ mass}$ is in fact a large negative
number at the beginning of the simulation.  Therefore, Figure \ref{fig:energy}
only displays the change during the simulation, $\Delta E_{\rm rest\ mass}$.
It is immediately evident that this is the principal energy source, as its
dependence is the direct converse of the energy deposition.

The final contribution to the total energy balance is the energy lost to
neutrinos during the neutronization process.  We define the cumulative
neutrino loss by
\begin{equation}
E_{\rm \nu-loss}(t) = \int_0^t\int_V \epsilon_\nu(t')\rho dV dt'\ ,
\end{equation}
where $\epsilon_\nu$ is the neutrino loss for the local conditions as
tabulated by \citet{SeitTownetal09} (see section \ref{sec:burning}).
The energy loss due to neutrinos is also shown in Figure \ref{fig:energy}, 
where it has been multiplied
by a factor of ten in order to be visible on this scale.  We observe that
there is no significant neutrino losses during the detonation phase,
indicating that the neutronization takes place exclusively during the
deflagration.

Adding all the energy terms together provides a useful check on the energy
conservation of our code.  The total is formed by $E_{\rm total} =E_{\rm
bind} + E_{\rm rest\ mass} + E_{\rm \nu-loss}$.  This sum again has a large
offset due to the binding energy of the unburned C-O-Ne material, so we only
display the change in total energy during the simulation.  Very good energy
conservation is observed, though there appears to be a small loss of energy
(barely perceptible in Figure \ref{fig:energy}) during heavy refinement and
derefinement, apparently related to $E_{\rm rest\ mass}$.  This loss may
be a unexpected interaction of interpolation with the representation of the
conserved quantities in the burning model.  The loss is small enough to not be a
concern, but will be addressed in the ongoing verification of our code
components.

\subsubsection{Transition to Expansion}

In order to obtain an accurate evaluation of the distribution of ejected
material in velocity, which is critical for spectral properties of the
explosion, it is necessary to continue the simulation until the ejecta has
reached approximate free expansion.  
The detonation has propagated throughout the interior of the star
by just after $t=1.6$~s as can be seen from Figure \ref{fig:sequence}.
Some nucleosynthesis takes place later as the NSE
freezes out, but the need for fine refinement of the energy generating
regions ends at this time since there are no propagating burning fronts.  As
a result, at 1.8~s in this simulation, the refinement of energy-generating
regions, including regions in which the propagation of RD front is still
taking place, is disabled.  Also the refinement of non-energy-generating
regions is coarsened from 16 to 32~km.  Ideally, de-refinement with further
expansion could take place automatically with appropriate detection of the
expansion of flow features.  This somewhat delicate balance was not
attempted in these calculations, and nearly the entire ejecta remains
refined at 32~km through the end of the calculation.

\begin{figure}
\plotone{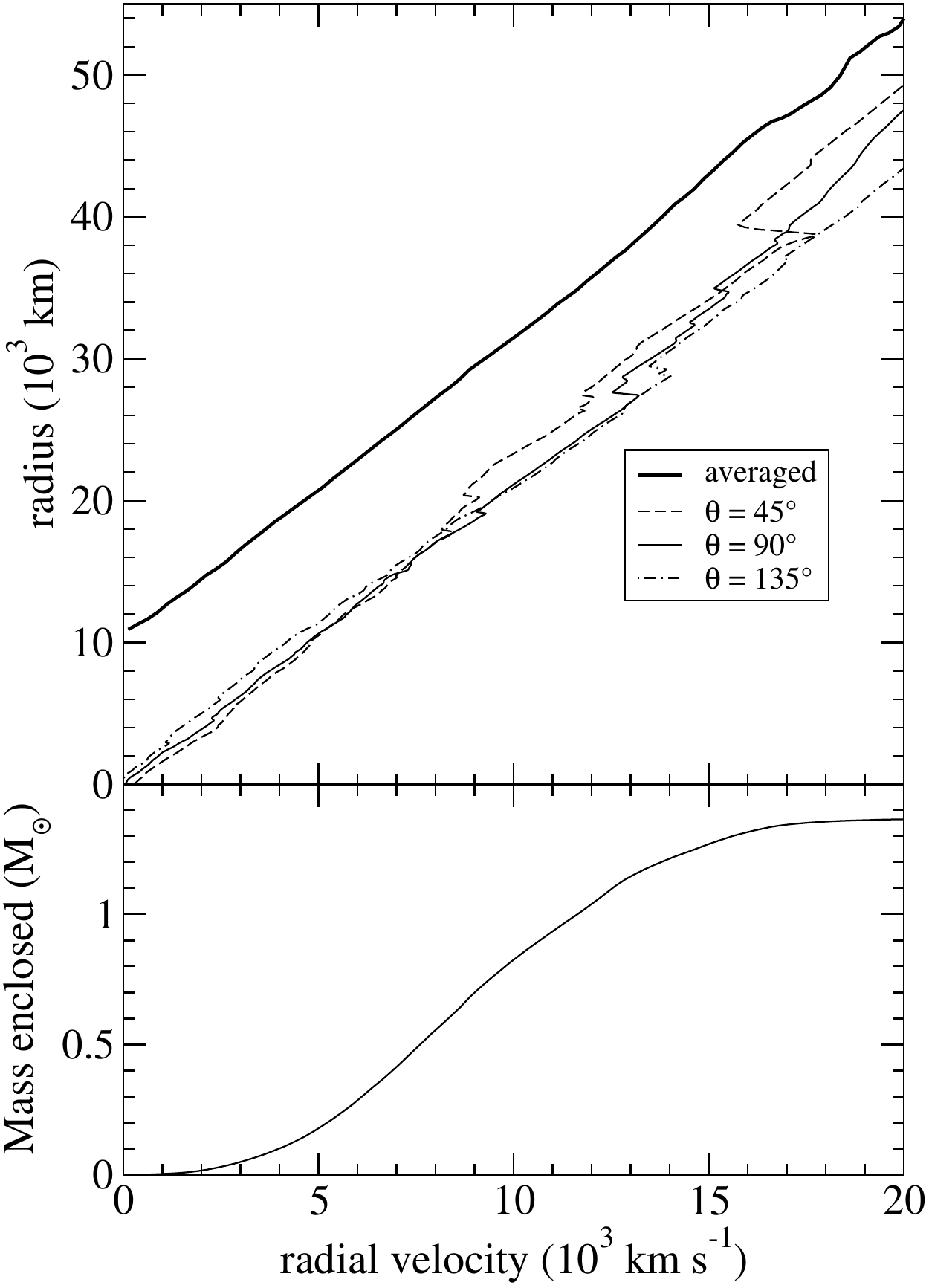}
%\plotone{r_M_profiles_ne_02_r2.pdf}
\caption{\label{fig:r_M_profile}
Run of radius and mass with radial velocity at $t=3.5$ s, demonstrating that
free expansion has been approximately attained.
}
\end{figure}
The calculation was halted as material began to flow off of the grid around
$t=3.5$~s.  The domain was chosen to be of such a size that this
time was late enough for the ejecta to be in approximate free expansion.
We take free expansion to be indicated by a linear relation between the
radial velocity and the radius.  Profiles of the radius against the radial
velocity of the outgoing ejecta are shown in Figure \ref{fig:r_M_profile}.
Shown are both line-out profiles at $\theta = 45$, 90, and 135$^\circ$, as
well as the averaged profile (offset by $+10$~Mm~s$^{-1}$ for clarity).
The average here is the mass weighted average radius for material moving at a
radial velocity within a bin of
0.2~km~s$^{-1}$.  While the
average profile has a fairly linear relation, there is some departure from
linear in the outer regions for certain directions in the ejecta.
Additionally, the radial velocity component completely dominates any motion
at this time.  Radial velocity contours are also shown on the final panel of
Figure \ref{fig:sequence} and are fairly symmetric, though there is a
noticeable degree of asymmetry in the abundance profiles.

\subsection{Explosion Yield and Ejecta Structure}

\begin{figure}
\plotone{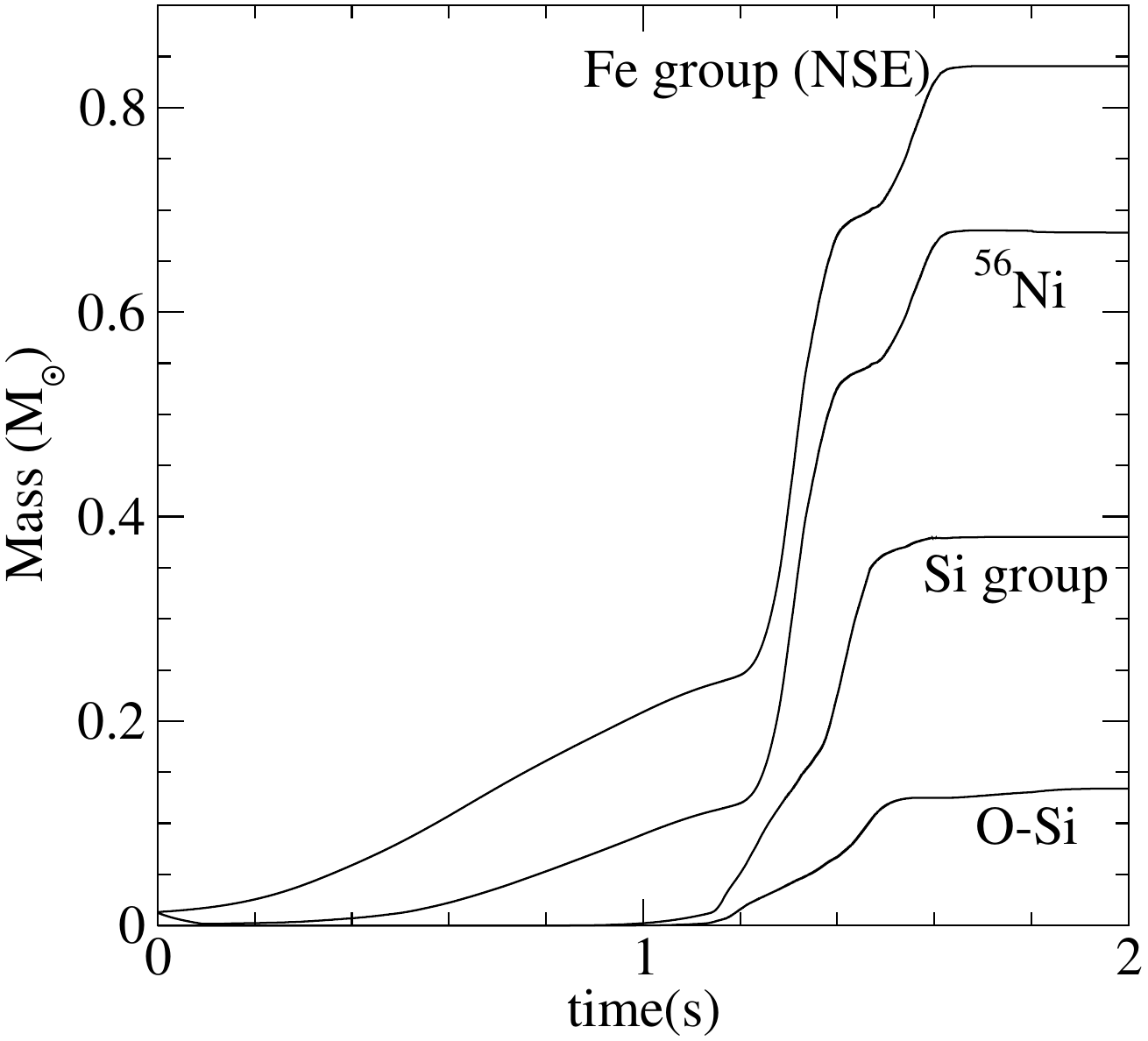}
%\plotone{masses_02_r2.pdf}
\caption{\label{fig:mass}
Masses of synthesized material in time during explosion, see text for
definitions.  Note that these are estimates of final outcome, and therefore
the relaxation of the NSE material dominantly Fe-group constituents
is not evident in this plot.
}
\end{figure}
The time history of the material produced in the incineration of the star is
shown in Figure \ref{fig:mass}.  In the interest of simplicity, we leave
tracer particle post-processing and detailed nucleosynthesis to future work
and instead just make use of the progress variables defined in the burning
model (see Section \ref{sec:burning}).  The ashes of the
first stage of burning are predominantly O and Si, with some other
intermediate mass elements (Ne, Mg), and its local mass fraction is given by
$X_{\rm O,Si}=\phi_{fa}-\phi_{aq}$.  Further burning produces a mixture of
various Si-group nuclides, predominantly $^{28}$Si, which begins in NSQE.  We
are concerned with the form that material will take in the outgoing ejecta,
after NSQE and NSE material have completely relaxed.  As such,  Figure
\ref{fig:mass} shows the final yield masses as simply Si- and Fe-group.  The
mass fraction of the Si-group material is given by $X_{\rm
Si-group}=\phi_{aq}-\phi_{qn}$.  Material in NSE is given by
$X_{\rm NSE} = \phi_{qn}$, and will consist, in the ejecta, of almost
entirely Fe-group elements.  These definitions are also used below to
characterize the abundance profile of the ejecta and allow the colors
displayed in figure \ref{fig:sequence} to be interpreted roughly in terms of
the nucleosynthetic products.

By tracking $Y_e$, we also have a local measure of the neutronization that
has occurred, and from this we can estimate how much of the NSE material will be
in the form of stable, neutron-rich Fe-group nuclides instead of $^{56}$Ni.
For $Y_e=0.5$, the ejected material is nearly pure $^{56}$Ni
\citep{Meaketal09}.  In order to estimate the local mass fraction of
$^{56}$Ni of a neutronized fluid element being ejected, we assume that the
next most abundant nuclides are an equal admixture of $^{54}$Fe and
$^{58}$Ni, as is observed in the tabulation of nucleosynthetic outcome with
$Y_e$ by \citet{Meaketal09}.  The local mass fraction of $^{56}$Ni is then
estimated by 
\begin{eqnarray}
 Y_{e,N} &=& [Y_e - (1-\phi_{qN})Y_{e,f}]/\phi_{qN}\nonumber\\
 X_{^{56}\rm Ni} &=&
 \max\left[\ \phi_{qN}\frac{Y_{e,N}-0.48212}{0.5-0.48212}\ , \ 0\ \right]\ .
\label{eq:xni56}
\end{eqnarray}
This estimated mass fraction is also used below in discussing abundance
profiles.

From Figure \ref{fig:mass}, nearly all of the intermediate mass elements
are produced during the detonation phase, with a small amount of Si group
material produced in the deflagration.  This latter is likely to be greater
for more expanded explosions with lower NSE and Ni yields.  The isolation of
the neutronization to the deflagration phase is also evident.  While
only half of the NSE material produced in the deflagration phase will be
ejected as $^{56}$Ni, nearly all of that produced in the detonation phase
will end up as $^{56}$Ni.  Note that since $X_{^{22}\rm Ne}=0.02$ for this
explosion, the maximum $X_{^{56}\rm Ni}$ even for non-neutronized material is
0.95.  Although it appears all the burning is complete by approximately 1.6
seconds, this is only true in terms of gross yields.  At that time much of
the Fe-group material will still be in NSE, and will relax as the star
expands.  The small glitch in the $^{56}$Ni curve at 1.8 seconds is due to the
cell-mixing upon zone de-refinement and the resulting change in our estimate of
the final $^{56}$Ni yield via equation (\ref{eq:xni56}).

\begin{figure}
\plotone{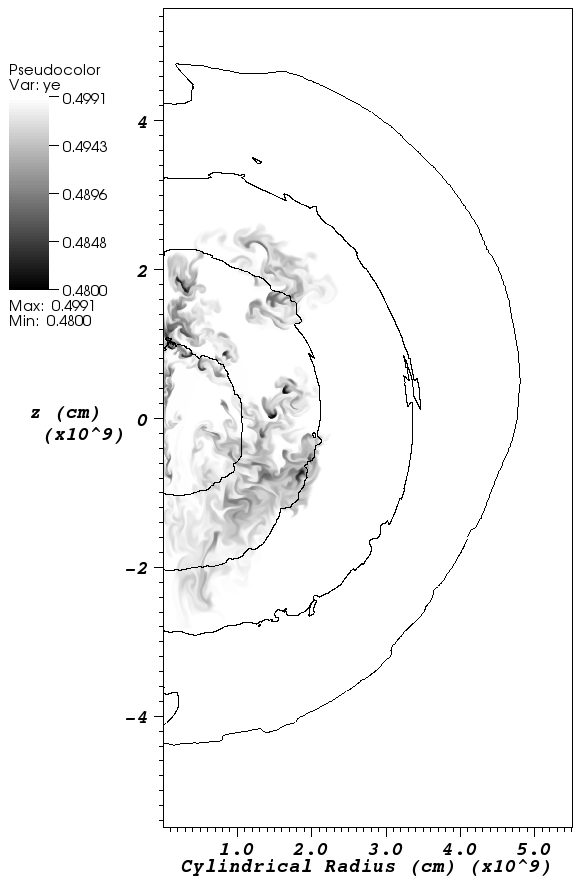}
%\plotone{ne_02_r2_ye_0175_clean_legend.png}
\caption{\label{fig:ye}
The electron molar fraction $Y_e$ at $t=3.5$~s after free expansion has been
essentially attained.  Contours indicated radial velocities of 5, 10, 15 and
$20\times 10^3$~km~s$^{-1}$.  View is the same as the last panel of Figure
\ref{fig:sequence}.
}
\end{figure}
The elemental ejecta structure is shown by the combination of the last panel
of Figure \ref{fig:sequence}, which shows radial velocity contours overlayed
on the color-coded gross yields as defined above, along with Figure
\ref{fig:ye} that gives the spatial distribution of $Y_e$ at the same moment
with the same radial velocity contours shown.  Material below
$Y_e\approx0.482$ is expected to have fairly little $^{56}$Ni, favoring more
neutron-rich nuclides instead.
\begin{figure}
\plotone{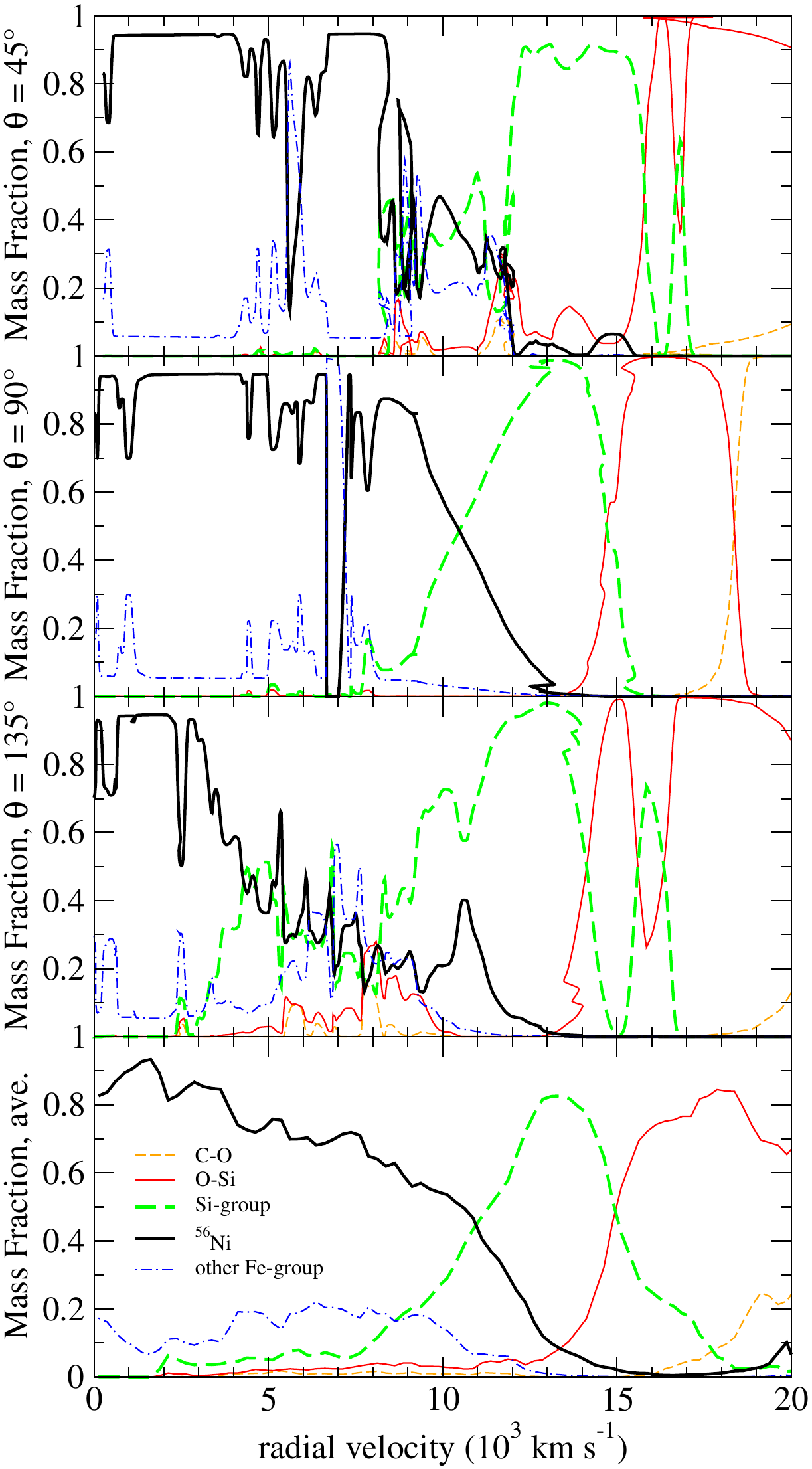}
%\plotone{X-v_profiles_ne_02_r2.pdf}
\caption{\label{fig:X_profile}
Run of burning products with radial velocity at $t=3.5$ s.
}
\end{figure}
For further detailed scrutiny, we have extracted the abundance pattern along
radial lines at several latitudes and the averaged abundance profile for
0.2~km~s$^{-1}$ bins, all shown in Figure \ref{fig:X_profile} in radial
velocity.  In this figure the $^{56}$Ni mass fraction is estimated from
equation (\ref{eq:xni56}).  In the case of the averaged profile this is done
before averaging.  The distribution of the mass in radial velocity is shown
in Figure \ref{fig:r_M_profile}, showing that very little mass is outside of
approximately 15,000~km~s$^{-1}$.
\begin{figure}
\plotone{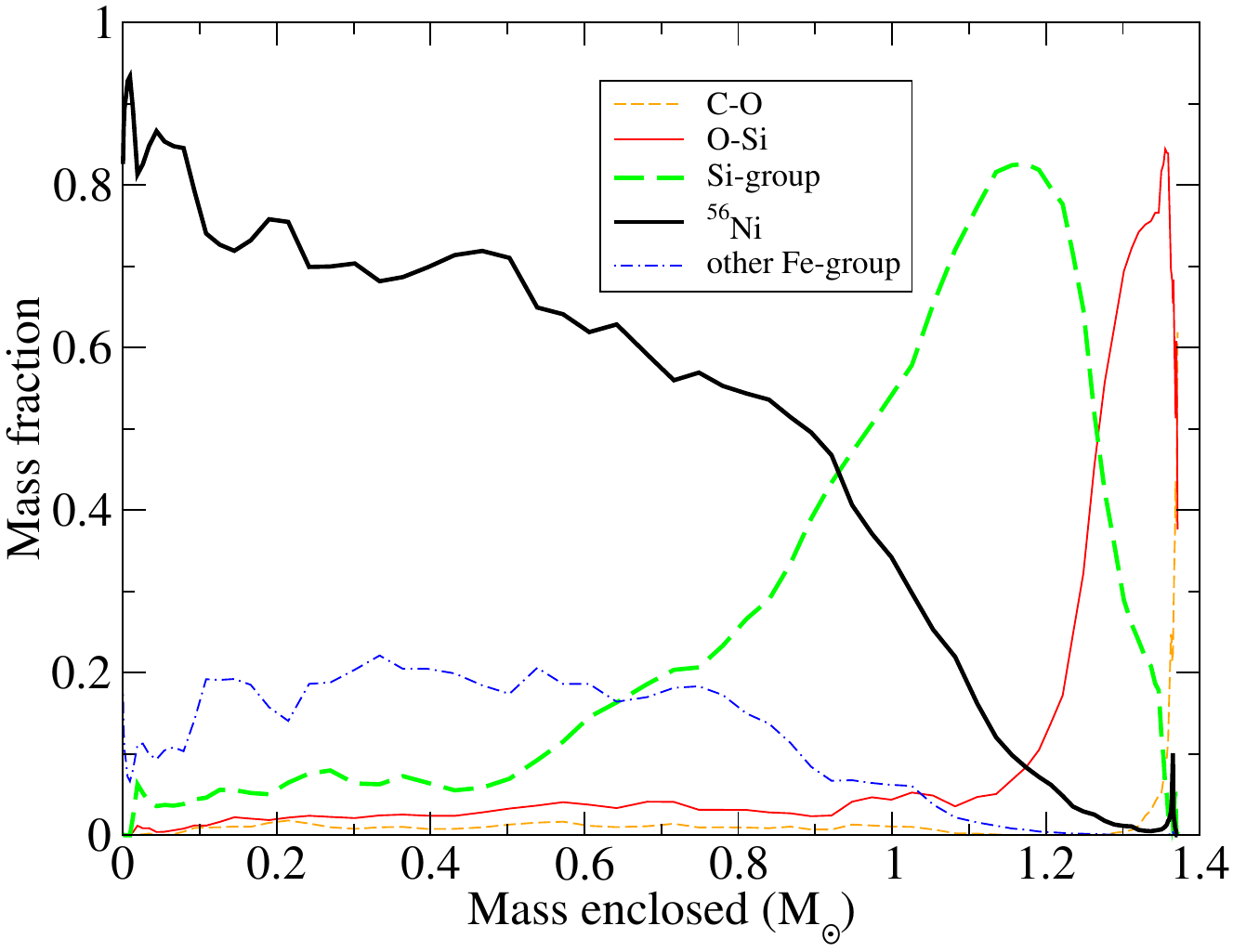}
%\plotone{X-M_profile_ne_02_r2.pdf}
\caption{\label{fig:X-M_profile}
Distribution of burning products with mass coordinate in ejecta.
}
\end{figure}
Finally the distribution of material in mass is shown in
Figure~\ref{fig:X-M_profile}, where the mass coordinate is defined as the mass
enclosed by consecutive radial velocity shells.

\subsection{Resolution Dependence}
\label{sec:resstudy}

Given the small scale of the expected flame surface structure 
with respect to the
grid scale in these simulations (4 km), and the very basic treatment
of turbulent acceleration of the burning, some dependence on resolution is
expected.  The important question is whether or not such resolution effects
are strong enough to preclude the use of suites of 4 km simulations from
being used to evaluate systematic effects.  We find that the apparent level
of uncertainty from the low resolution is acceptable when considering
particular metrics, but that there are important dependencies on resolution
that we hope can be resolved as treatments of subgrid flame structure and
their usage becomes more advanced.

\begin{figure}
\plotone{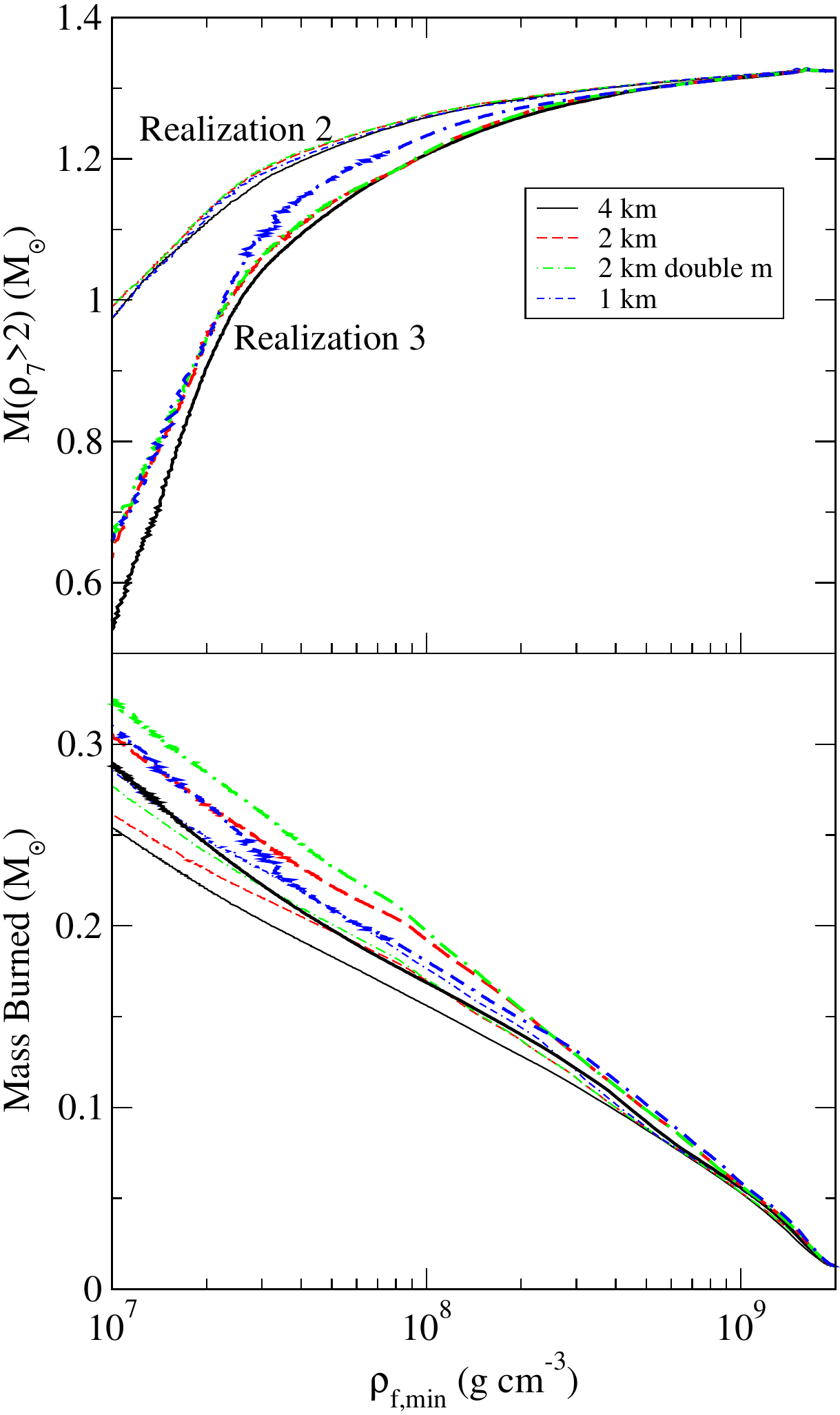}
%\plotone{res_compare_Mhi_Mburn-dens.pdf}
\caption{\label{fig:res_Mhi_Mburn}
Investigation of the dependence of results on resolution.  The
top panel shows the mass above the density threshold $2\times
10^7$~g~cm$^{-3}$, found in section \ref{sec:population} to correlate well
with total NSE yield.  The curves are plotted against 
the minimum density of fuel
being currently burned by the flame, $\rho_{f,\rm min}$, such that the
evolution follows the curve from right to left as the flame rises and the
star expands.  The bottom panel shows the total mass burned, essentially all
of which is NSE material.  While the value of $M(\rho_7>2)$ when $\rho_{f,\rm
min}=\rho_{\rm DDT}$ appears fairly robust with resolution, providing some
confidence in its usage to study systematic effects in section
\ref{sec:systematics},  the total burned mass shows more resolution
dependence.
}
\end{figure}

The metric used in our study of systematics in section \ref{sec:ne22systematic}
below is the mass at high density at the DDT time.  This time is defined by
the moment at which the lowest density of fresh fuel being encountered by the
flame, $\rho_{f,\rm min}$, falls below $\rho_{\rm DDT}$, which we are assuming
here is $10^7$~g~cm$^{-3}$.  It was found in section \ref{sec:population} that
a good density threshold to use to define the mass at high density is
$2\times10^7$~g~cm$^{-3}$.  Thus it is useful to look at how the evolution of
$M(\rho_7>2)$ as the star expands, and the flame rises, depends on resolution.
The top panel of Figure~\ref{fig:res_Mhi_Mburn} displays the dependence of
$M(\rho_7>2)$ on $\rho_{f,\rm min}$.  Evolution in this figure moves from
right to left as the flame rises through the star.  Two different
realizations from the sample developed in section~\ref{sec:population} are
shown.  One is that described in detail in this section (realization 2), 
and the other is a
more extreme case that produces much less NSE material (realization 3).
In addition to several resolutions, a case was run with the value of the
parameter $m$ that appears in the buoyancy-compensation prescription for the
flame speed (see section \ref{sec:flamespeed}) doubled.  By simultaneously
doubling this parameter and halving the resolution, the actual value of the
prescribed flame speed stays the same.

Our metric for measuring the expansion of the star, $M(\rho_7>2,t_{\rm
DDT})$, shows a fairly low sensitivity to resolution for these cases,
especially for realization 2.  There is also fairly little sensitivity to the
value of the parameter $m$.  This insensitivity is likely due to the fact
that the two competing processes at work, the pulsational expansion of the star and the
rise of burned plumes through the star, are not overly sensitive to the
poorly resolved flame surface.  The small-scale flame structure is not
important for the selection of the dominant plume because this is largely
determined by the initial condition (see also discussion in section
\ref{sec:population}).  Also, due to the delay in manifestation of the
turbulence from the Rayleigh-Taylor instability of burned plumes, the
additional flame surface area and concomitant energy deposition may come after
the critical phase for the launch of the pulsational expansion of the star.
This delay in energy release might lessen the impact of uncertainty 
in the burning rate during the
intermediate stages of the deflagration, the time when it is most difficult
to model accurately.  However, this situation highlights the necessity for a careful
understanding of the very early portion of the deflagration phase and the
interaction of the flame surface with the turbulent convection field arising
from the simmering phase.

There is a stronger and more systematic dependence of the total burned mass
on resolution, as shown by the bottom panel of
Figure~\ref{fig:res_Mhi_Mburn}.  However, the scatter in the consecutive
resolution cases performed makes them difficult to interpret clearly.  In an
observational sense, the resolution dependence might be closely related to
the weakness of the neutron-enriched core observed in the yields from
realization 2, as discussed below.  The additional enhancement of burning due to
turbulence should be most effective in the inner regions of the star where
the intersecting wakes of rising plumes will lead to strong turbulent
shearing of the flame surface.  This boost may burn core material early enough in
the expansion to enable this material to undergo significant electron
captures.  The two-dimensional simulations performed here (in which there is no
physically realistic turbulence cascade) and the lack of an explicit
treatment of turbulence-flame interaction make it inappropriate for this to
be addressed in the current study.  We will simply leave the degree to which
turbulence influences the neutronization of material an open question, which is not
expected to depend on the $^{22}$Ne content under study here due to the
predominance of turbulence in setting the fuel consumption rate under these
conditions.

\subsection{Brief Comparison with Observed Properties of SN~Ia}

As seen from Figures \ref{fig:sequence}, \ref{fig:ye}, \ref{fig:X_profile}, and
\ref{fig:X-M_profile}, our explosion reproduces fairly well the
observed radial velocity structure of the intermediate mass elements ejected
in the SNe~Ia \citep{Fili97}.  We will leave a fully detailed comparison
until full abundances can be obtained from post-processing tracer particles,
but many important features warrant mentioning here.  There is a prominence of Si in
abundance for the velocity range of roughly 10 to 15~Mm~s$^{-1}$,
accounting for nearly $0.4M_\odot$ of the final ejecta, with some Si-group
material extending down as far as 8~Mm~s$^{-1}$.  This is broadly consistent
with the spectral evolution of SNe~Ia, though possibly 1-2~Mm~s$^{-1}$ high
in velocity compared to the spectral evolution of the most standard cases
like 1994D \citep{Pataetal96}.  Since the mass distribution (Figure
\ref{fig:r_M_profile}) is weighted toward lower velocities, conclusive
comparison will require radiative transfer calculations.  Also, this is only a
single case of many possible, which may have Si at lower velocities.  The
ejecta is fairly symmetric, with a slight asymmetry arising from the location
and timing of the detonation sites.  While the degree of asymmetry is likely
not too unrealistic, the pattern imprinted may be a bit different in a
three-dimensional treatment.  It does appear that the velocity extent of Si
features near the photosphere would be viewing-angle-dependent.  Note that
the jet-like columns of Fe-group material on the symmetry axes are
unrealistic, but appear to have a fairly small contribution to the averaged
profile shown in the last panel of Figure \ref{fig:X_profile}.  Detailed
radiative transfer calculations will be required to determine if the
patchiness of the border between the Si and Fe-group material provides a good
match to polarimetric observations \citep[e.g.][]{Wangetal03}.

The distribution of neutron-rich material demonstrated in Figures
\ref{fig:ye} and \ref{fig:X_profile} compares less well with the distribution
inferred from observations \citep{Stehetal05,Mazzetal08}.  On the positive
side,  the impurity created by the distribution of deflagration material at
velocities around 3-8~Mm~s$^{-1}$ dilutes the pure $^{56}$Ni to a degree that
provides a reasonable agreement with observations.  However, notably absent
is the predominance of stable Fe at low velocities, $\lesssim 3$~Mm~s$^{-1}$,
which is inferred from observations of 2002bo \citep{Stehetal05} and 2004eo
\citep{Mazzetal08}.  While the treatment of the ejecta evolution leading to
this inferred abundance is approximate, it seems unlikely that such a stark
difference could be reconciled without a change in the model.  The most
apparent reason for this shortcoming of the model is the flame treatment
being utilized here.  The buoyancy-compensation form of \cite{Khok95}, which
we are using, does not account for interaction of the turbulent wakes with
other burning fronts.  Thus as plumes rise out of the core, the turbulence
they leave behind does not lead to the enhancement of flame spreading that it
should.  This shortcoming will be addressed in future work, with more
sophisticated treatments of the turbulence-flame interaction
\citep{Colietal00,Schmetal06a}.  Investigation of this effect should be
undertaken with attention to the detailed nuclear products, as extensive
central burning early in the deflagration phase can lead to
"over"-neutronization \citep[e.g.][]{Bracetal00}.

Many SNe~Ia have been observed with high-velocity Ca features
\citep{Mazzetal05}.  This material is generally produced during the
transition from NSQE to NSE~\citep[e.g.][]{Nomoetal84} and thus should appear
at the border between Si-group (green) and NSE (black) regions
in Figure \ref{fig:sequence}.  In this simulation, these
yields are restricted to velocities $\lesssim 13,000$~km~s$^{-1}$, where the
main component also lies in observed supernovae.  Some features, however, do
indicate that this material might be produced at higher velocities by
intersecting detonation fronts.
The case of the symmetry axes, while an artifact here, suggests that points
where 3 detonations spreading through the star meet might be good candidate
sites for the formation of such high-velocity features.  This is reinforced
by features seen in some detonation interactions in other realizations (see
the comparison of ignition conditions in Figure \ref{fig:ddt_compare}).
Realistic tests will await DDT calculations in three dimensions.  Additionally,
we have assumed that detonations are ignited when the first flame front
passes through the DDT density.  This may be too stringent a condition, and
some number of deflagration plumes may extend well above this transition
region (which also
corresponds closely to the division between Si- and NSE-rich ashes) before
the detonation takes place.

%%%%%%%%%%%%%%%%%%%%%%%%%%%%%%%%%%%%%%%%%%%%%%%%%%%%%%%%%%%%%%%%%%
\section{Systematic Effects of $^{22}$Ne Arising from Dynamics}
\label{sec:ne22systematic}

We have implemented the current study as an attempt to determine whether the
presence of $^{22}$Ne has a significant systematic effect on the expansion
state of the star at the time of first detonation.  In light of the
relation between the expansion and final yields discussed in Section
\ref{sec:population},
this is a likely candidate for an effect of $^{22}$Ne that can
compete with that expected due to the simple overall neutron excess
highlighted by \cite{timmes.brown.ea:variations}.
Our focus on the degree of expansion, quantified by $M(\rho_7>2,t_{\rm DDT})$, in such
specificity is motivated by a desire to separate the contribution of
$^{22}$Ne to the overall nucleosynthesis into a dynamical component
distinguishable from the gross neutron excess.  The use of the outcome of the
deflagration phase alone also allows many more cases to be run at the same
computational cost.  This is very important given the high variance in
the outcome of the supernovae.  Averaging over many realizations is
necessary to obtain uncertainties that are small compared to the expected
effect of neutron excess.  Additionally, targeting our study so
specifically gives more power for tertiary conclusions about the impact of
other effects that might modify the propagation speed of the burning front.

\begin{figure}
\plotone{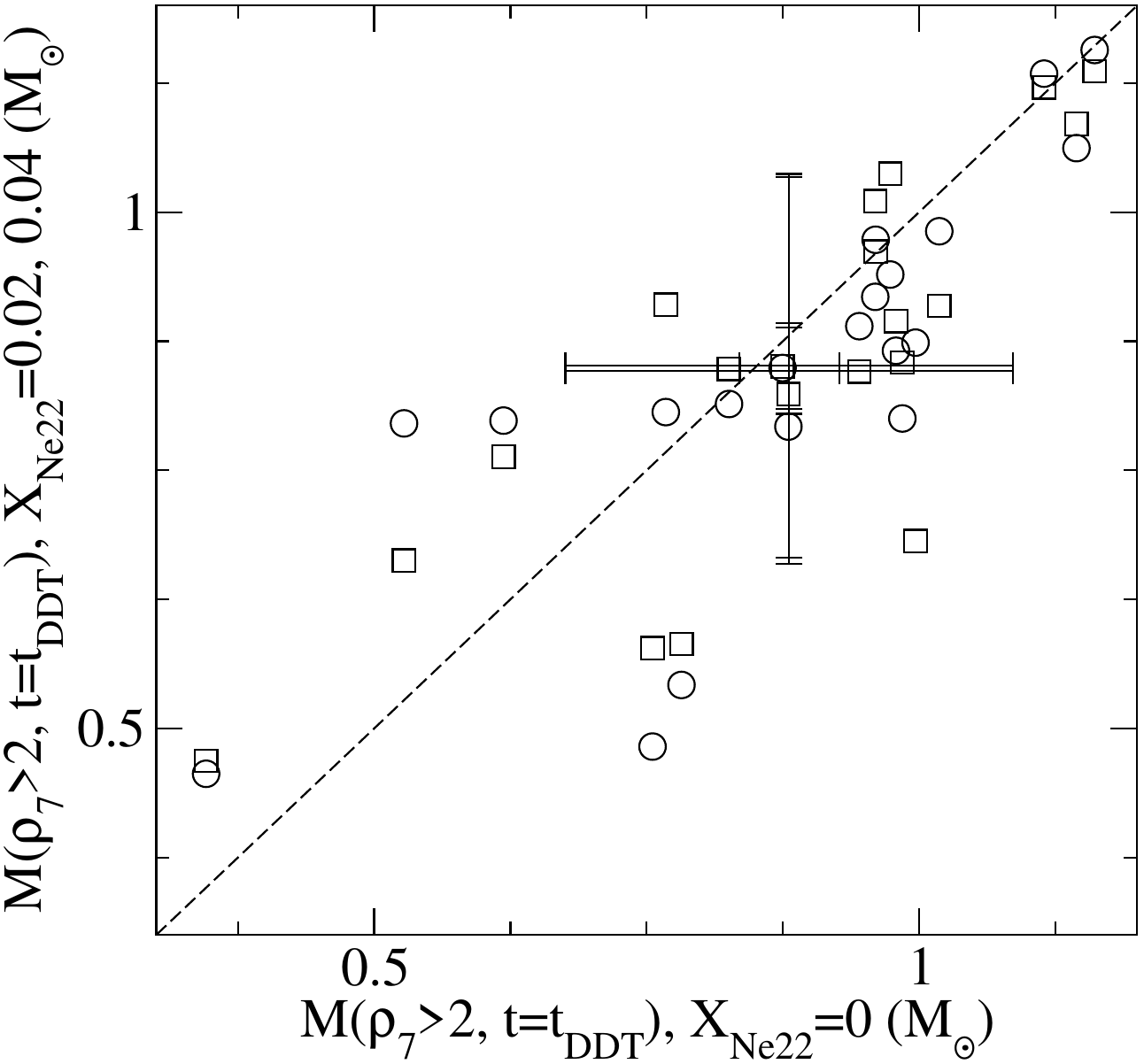}
%\plotone{Mhi_00_02_04.pdf}
\caption{\label{fig:Mhi_00_02}
Relation of degree of pre-expansion for cases with
$X_{^{22}\textrm{Ne}}=0.02$ (circles) and $0.04$ (squares), to a case with no
$^{22}$Ne.
Degree of pre-expansion is indicated by the mass above a density of $2\times
10^7$ ~g~cm$^{-3}$ when the first detonation initiates, which is an indicator
of how much Fe-group material will be ejected in the supernova.
The point with vertical and horizontal error bars indicates the average,
sample standard deviation (outer error bars) and standard deviation of the
mean (inner error bars).  The sample with $X_{^{22}\textrm{Ne}}=0.04$ has a
slightly higher average.  There is not a significant difference between the
average expansion in the three cases.
}
\end{figure}

Variation due to $^{22}$Ne was explored by simulating the deflagration phase
for 20 realizations with $X_{^{22}\rm Ne}=0$, 0.02, and 0.04.  As discussed in
section \ref{sec:compositionsystematics}, the carbon abundance was kept fixed at $0.5$ by
mass as was the central density of the initial WD.  This results in the WD
being slightly less massive with a higher $^{22}$Ne content.  Our WD, as
mapped on the initial grid, is $2.7417$, $2.7314$ or $2.7211\times
10^{33}$~g for $X_{^{22}\rm Ne}=0$, 0.02, and 0.04 respectively.
Samples with different $^{22}$Ne content are compared in Figure
\ref{fig:Mhi_00_02}, where $M(\rho_7>2,t_{\rm DDT})$ for the $^{22}$Ne
enhanced cases is plotted against the same outcome for no $^{22}$Ne.  The
dashed line indicates a 1-1 relation.  Systematic effects should arise as a
departure from this relation on average.  The two nearly coincident points
with error bars indicate the averages of the samples with different $^{22}$Ne
abundances.  The outer error bars indicate the standard deviation and the
inner the standard deviation of the mean.  The sample with $X_{^{22}\rm
Ne}=0.04$ has a slightly higher average of 0.851$M_\odot$ compared to
$0.846M_\odot$ for $X_{^{22}\rm Ne}=0.02$, and both of these are somewhat
lower than the $0.881M_\odot$ for no $\Ne$.  Standard deviations of the mean
are 0.042$M_\odot$ for both of the nonzero $\Ne$ cases and 0.046$M_\odot$ for
the zero $\Ne$.

The overlap of the inner error bars with the 1-1 relation indicates that
there is no statistically significant impact of $^{22}$Ne content on the
expansion prior to DDT over this range of mass fractions. On the basis of
neutron excess \citep{timmes.brown.ea:variations}, the decrease in the
$^{56}$Ni mass between $\Ne$ fractions of 0 and 0.04 is $\simeq 10$\%.  Our
results preclude a difference in the overall NSE production that could
compete with this change, finding it to be $\lesssim 5\%$.
The more
physically motivated interval between 0.02, appropriate for a zero
metallicity progenitor, and 0.04, appropriate for an approximately solar
metallicity progenitor \citep{PiroBild07,Chametal07}, shows a very small
difference.  Note that changes in $X_{^{22}\rm Ne}$ by 0.02 can lead to changes
in individual cases by as much as 0.2$M_\odot$, but that this effect is
washed out by averaging over a variety of ignition conditions.  This implies
that the important controlling processes are large-scale plume rise and
stellar expansion, and that these are not influenced in a systematic way by
modest changes in the flame propagation like those introduced by $^{22}$Ne.
This, in fact, bodes well for the robustness of simulations in the DDT
scenario, as many features of the explosion don't depend on accurate
treatment of physical phenomena at all scales.  However, far more
investigation is necessary before such a robustness can be actually claimed.

Returning to the cases shown in Figure~\ref{fig:ddt_compare}, the addition of
$^{22}$Ne does not change the major morphology of the deflagration.
Generally the same dominant plumes are observed at both $^{22}$Ne fractions.
However, the substructure of the rising plumes can undergo some fairly
extensive modification, which can also influence the timing of their rise and
therefore the detonation.  The modification likely arises from influence of the flame
speed on the early evolution of structure on the smallest scales of the
initial flame surface.  The absence of a systematic effect implies that 
such modification 
is equally effective at suppressing dominant plumes and enhancing
sub-dominant ones.  This supposition is demonstrated anecdotally in
Figure~\ref{fig:ddt_compare}, where, for example, in realization 5, enhancing
the $^{22}$Ne enhances one secondary plume and suppresses another.  This
effect 
might be a manifestation of the tendency of R-T to cause perturbations to
grow at the expense of others, such that selection of the dominant
perturbation is less important than the dynamics of the dominant perturbation
once one "wins".

There is still an important way for $^{22}$Ne to influence the dynamics of
the explosion not addressed here.  This study treated a sample of ignition
conditions with fixed harmonic content.  If, however, the change in the flame
speed due to $^{22}$Ne has a significant impact during the earliest stages of
the deflagration, it may change the effective harmonic content of an
abstracted initial condition as implemented here.  However, the apparent lack
of sensitivity, on average, to changes in the small scale structure of
deflagration morphology argue against this being important.  Thus, the
important features of the ignition condition may be set by the large scale
flows in the convective core and the relative position of the initial spark.
This subject deserves attention as simulations of the early part of the
deflagration phase are undertaken.

Although we have held other parameters fixed for this study, as discussed in
section \ref{sec:compositionsystematics}, the $^{22}$Ne composition is not
the only parameter which may vary with the metallicity of the parent stellar
population.  Notably the central carbon fraction and ignition density are
expected to depend on both the metallicity and the main-sequence mass of the
progenitor star.  The ignition density may additionally depend on the
accretion history of the progenitor WD.  A full accounting of the
systematic dependence of theoretical SNe~Ia will therefore have to await evaluation of
these further parameters, enabled by the framework presented here.

%%%%%%%%%%%%%%%%%%%%%%%%%%%%%%%%%%%%%%%%%%%%%%%%%%%%%%%%%%%%%%%%%%
\section{Conclusions and Future Work}
\label{sec:conclusions}

We have extended a deflagration ignition condition of constrained asymmetry,
with no harmonic components less than $\ell=12$ in the initial flame surface,
to develop a well-defined random sample, and use this to construct a basic
framework for formal study of systematic effects
in SN~Ia.  Our theoretical sample has a $^{56}$Ni mass average and standard
deviation of 0.7 and 0.11$M_\odot$ (0.62 and 0.11$M_\odot$) for a $\Ne$
fraction of 0 (0.02) by mass, and $^{56}$Ni yield ranging between 0.45 and
0.8$M_\odot$.  This sample compares well with that of observed SN~Ia
\citep[e.g.][]{howelletal+09}, and should be somewhat tunable by varying the
harmonic content of the initial condition and/or the DDT density.  Our
framework uses this sample to enable statistically well-defined
studies of both physical and theoretical parameters of the SN~Ia explosion
simulation.  A great virtue of this methodology is that it includes the
inherent diversity of SN~Ia outcomes and the real-world challenges that
brings for the study of systematic effects.  The framework is expected to
undergo refinement and extension to make the simulations of the SN~Ia
increasingly realistic.  Important studies for these improvements include
improved turbulence-flame interaction models and a more careful
characterization of the relation between spectral content in the initial
condition and simulation outcome.

The outcome of a typical two-dimensional simulation of an SN~Ia from our
sample with a DDT density of $10^7$~g~cm$^{-3}$ was presented in detail and
compared to observations.
This model explosion provides a reasonable reproduction of many features of
observed SN~Ia.  Intermediate mass elements are dominant at velocities above
10~Mm~s$^{-1}$, and a mixture of neutron-enriched Fe-group and $^{56}$Ni
dominates at lower velocities.  Material is fairly well constrained to layers
in velocity, with a modest degree of overall asymmetry. There is also some
large-scale mixing at the border between Si-group and Fe-group dominated
material in the velocity range $\sim 8$-11~Mm~s$^{-1}$, due to tops of the
highest plumes during the deflagration phase.
For this study we have relied upon the parameterized burning model used in
the hydrodynamic simulation in order to give a gross measure of the
nucleosynthetic outcome.  Future work will proceed with post-processing
Lagrangian particle tracks, yielding full nucleosynthetic information.

Besides the many features that match observations well, it appears that our
simple prescription for turbulence-flame interaction, which only compensates
for adverse effects of buoyancy on the integrity of the burning front, causes
an underproduction of stable Fe-group elements at low velocities ($\lesssim
3$~Mm~s$^{-1}$).  Improved treatments that have been proposed in the
literature \citep{Colietal00,Schmetal06a} will be included in future work,
allowing a direct comparison.  Notably, however, we do not find a strong
resolution dependence for the competition between the stellar expansion and
plume rise that sets the star's density structure at detonation and,
therefore, the overall yield of NSE and intermediate mass material.  This
result
indicates that our prescription is sufficient for its main intended purpose
of allowing an approximate capture of the dynamics of large-scale plumes.

Although we briefly reviewed the relevant dependencies that must be considered
in order to evaluate the full systematic effect of stellar population
metallicity (Section \ref{sec:compositionsystematics}), for the first application of the
framework we focus on a question of more well-defined scope.  We fix
the carbon content of the white dwarf to half by mass, and assume the
same accretion history, so that the ignition density is the same.  We also
assume the DDT density has a single value, $10^7$~g~cm$^{-3}$.  These
assumptions make our study specifically tailored to understand how dynamical
effects might change the simple relationship between $\Ne$ content and
$^{56}$Ni production presented by \citet{timmes.brown.ea:variations}.

\begin{figure}
\plotone{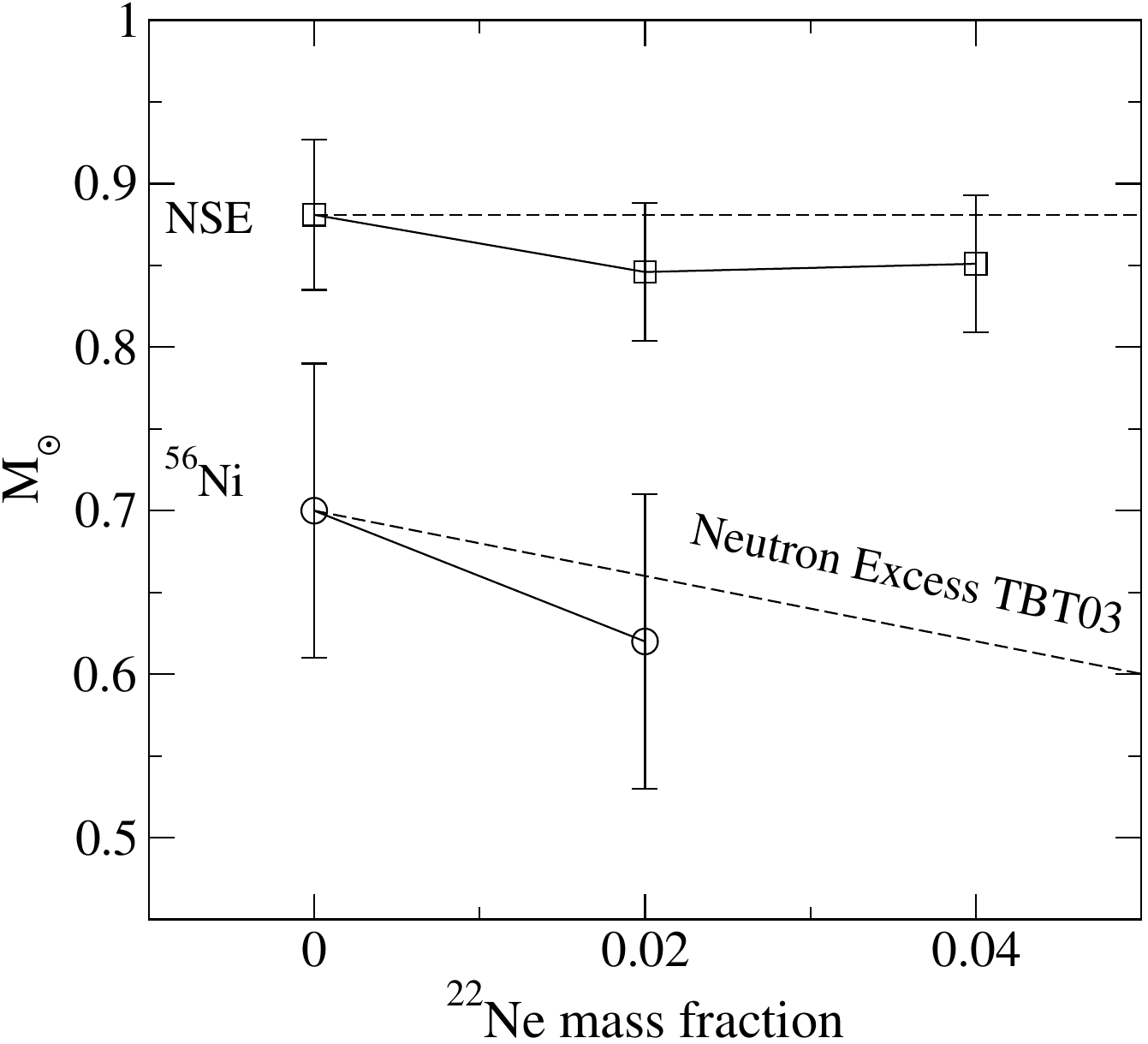}
%\plotone{ne_trend.pdf}
\caption{\label{fig:ne_trend}
Comparison of our simulation results and the trends predicted from variation
in inherited neutron excess alone.
\citet{timmes.brown.ea:variations} predicts that for constant NSE mass
produced in the SN~Ia (upper dashed line), the $^{56}$Ni mass produced
decreases linearly with progenitor $^{22}$Ne content (lower dashed line).
Our results from 20 simulations of the SN~Ia deflagration phase at each
abundance indicate that the average NSE masses (squares) are consistent with
being independent of $^{22}$Ne content, and that the trend of $^{56}$Ni
(circles) is therefore consistent with that predicted by
\citet{timmes.brown.ea:variations}.  Only a subset of simulations were run
far enough to obtain final $^{56}$Ni yields.
}
\end{figure}

We proceed by calibrating an indicator of the degree of expansion that takes
place prior to the detonation ignition, which is chosen to be representative
of the total amount of NSE material produced in the explosion.  We find that
the mass above a density of $2\times 10^7$~g~cm$^{-3}$ when the DDT takes
place, $M(\rho_7>2,t_{\rm DDT})$, provides a good indicator of the final
NSE mass.  This both provides a very direct probe of the dynamical
contribution of $\Ne$ and can be evaluated efficiently for many realizations
in the theoretical sample, giving good statistics.
By averaging over 20 realizations of initial condition, we find that
$M(\rho_7>2,t_{\rm DDT})$ is 0.881, 0.846 and 0.851~$M_\odot$ for
$X_{\Ne}=0$, 0.02, and 0.04 respectively all with a standard deviations of
the mean of approximately 0.04$M_\odot$.  These results are shown in Figure
\ref{fig:ne_trend} as estimated NSE masses (squares).
There is no statistically significant dependence of the star's expansion, and
therefore the total NSE mass produced, on the $\Ne$ content.  In individual
cases, however, $M(\rho_7>2,t_{\rm DDT})$ can vary by as much as 0.2$M_\odot$
due to a change in $X_{\Ne}$ of 0.02.  Further, any possible effect is
smaller than the 10\% reduction in $^{56}$Ni mass predicted by
\citet{timmes.brown.ea:variations} on the basis of neutron excess for this
range of $\Ne$ fraction assuming constant NSE mass.  Their prediction, which
assumes that NSE mass is independent of $\Ne$ content, is shown by dashed
lines in Figure \ref{fig:ne_trend}.  Also shown is the average $^{56}$Ni
mass yields from the 5 simulations at each of $X_{\Ne}=0$ and 0.02.  For
these, the error bars are obtained using the standard deviation of the full
sample, as this small subset underestimates the variance.  These are
consistent with the reduction due to neutron excess, although the statistical
uncertainty is fairly large.

Our studies point to the morphology of the ignition condition as being the
dominant dynamical driver of the $^{56}$Ni yield of the explosion.  In two
dimensions this is manifested by whether the flow is dominated by an
equatorial or polar plume, but there should be a analogous morphological
criteria in three dimensions.  This points to the importance of the very early
deflagration phase, during which the region taken here as already burned is
formed.  The interaction of flame propagation and turbulence in this region
will set the harmonic content of the initial condition for a study such as
this one.  This provides a further opportunity for $\Ne$ to be important
through its effect on the flame propagation speed.

Finally it is important to keep in mind that the effect of $\Ne$ on both the
propagation speed and width of the flame is expected to change the DDT
density \citep{Chametal07} even if the dynamics is not affected on average.
This will be addressed in future work.

\acknowledgements

DMT is the Bart J. Bok fellow at Steward Observatory, The University of
Arizona.  This work was supported by the Department of Energy through grants
DE-FG02-07ER41516, DE-FG02-08ER41570, and DE-FG02-08ER41565, by NASA through 
grant 08-ATFP08-0062, and by the National Science Foundation through the 
Joint Institute for Nuclear Astrophysics
(JINA) under grant PHY 02-16783 and by grant AST 05-07456.  ACC also 
acknowledges support from the 
Department of Energy under grant DE-FG02-87ER40317. The authors acknowledge 
the hospitality of the Kavli Institute for Theoretical Physics, which is 
supported by the NSF under grant PHY05-51164, during the program ``Accretion 
and Explosion: the Astrophysics of Degenerate Stars.''  The software used in 
this work was in part developed by the DOE-supported ASC/Alliances Center 
for Astrophysical Thermonuclear Flashes at the University of Chicago. We 
thank Nathan Hearn for making his QuickFlash analysis tools publicly 
available at http://quickflash.sourceforge.net.  We also thank the anonymous
referee for a careful reading of the manuscript and constructive
comments.  Simulations presented in 
this work were run at the High Performance Computing Center at Michigan State 
University. This research utilized resources at the New York Center for 
Computational Sciences at Stony Brook University/Brookhaven National 
Laboratory which is supported by the U.S. Department of Energy under Contract 
No. DE-AC02-98CH10886 and by the State of New York.  

\bibliography{master,timmes_master,townsley_master}

\end{document}